\pdfoutput=1
\documentclass[conference]{IEEEtran}
\IEEEoverridecommandlockouts
\usepackage{cite}
\usepackage{amsmath,amssymb,amsfonts}
\usepackage{graphicx}
\usepackage{textcomp}
\usepackage{xcolor}
\usepackage{enumitem}
\usepackage{subfig}
\usepackage{url}
\usepackage[ruled, lined, linesnumbered, commentsnumbered, longend]{algorithm2e}
\usepackage{booktabs}
\usepackage{multirow}
\usepackage{makecell}
\usepackage{float}

\usepackage{listings}
\definecolor{codegreen}{rgb}{0,0.6,0}
\definecolor{codegray}{rgb}{0.5,0.5,0.5}
\definecolor{codepurple}{rgb}{0.58,0,0.82}
\definecolor{backcolour}{rgb}{0.98,0.98,0.98}

\lstdefinestyle{mystyle}{
    backgroundcolor=\color{backcolour},   
    commentstyle=\color{codegreen},
    keywordstyle=\color{magenta},
    numberstyle=\tiny\color{codegray},
    stringstyle=\color{codepurple},
    basicstyle=\ttfamily\footnotesize,
    breakatwhitespace=false,         
    breaklines=true,                 
    captionpos=b,                    
    keepspaces=true,                 
    showspaces=false,                
    showstringspaces=false,
    showtabs=false,                  
    tabsize=2
}
\lstdefinelanguage{p4}
{morekeywords={table, action, Register, RegisterAction}, 
sensitive=false, 
}

\def\BibTeX{{\rm B\kern-.05em{\sc i\kern-.025em b}\kern-.08em
    T\kern-.1667em\lower.7ex\hbox{E}\kern-.125emX}}
\begin{document}
\bstctlcite{IEEEexample:BSTcontrol}
\title{Genos: General In-Network Unsupervised\\ Intrusion Detection by Rule Extraction
}


\author{%
  Ruoyu Li$^{\mathsection\dagger}$, 
  Qing Li\thanks{Corresponding author: Qing Li (liq@pcl.ac.cn).}$^\dagger$, 
  Yu Zhang$^\mathsection$,
  Dan Zhao$^\dagger$,
  Xi Xiao$^\natural$,
  Yong Jiang$^{\natural\dagger}$\\
  $^\mathsection$Tsinghua University, China; $^\dagger$Peng Cheng Laboratory, China\\
  $^\natural$Tsinghua Shenzhen International Graduate School, China\\
  \texttt{\{liry19,yu-zhang23\}@mails.tsinghua.edu.cn};\\
  \texttt{\{liq,zhaod01\}@pcl.ac.cn}; \texttt{\{jiangy,xiaox\}@sz.tsinghua.edu.cn}
}

\maketitle

\begin{abstract}
Anomaly-based network intrusion detection systems (A-NIDS) use unsupervised models to detect unforeseen attacks.
However, existing A-NIDS solutions suffer from low throughput, lack of interpretability, and high maintenance costs. Recent in-network intelligence (INI) exploits programmable switches to offer line-rate deployment of NIDS. Nevertheless, current in-network NIDS are either model-specific or only apply to supervised models.
In this paper, we propose Genos, a general in-network framework for unsupervised A-NIDS by rule extraction, which consists of a Model Compiler, a Model Interpreter, and a Model Debugger.
Specifically, observing benign data are multimodal and usually located in multiple subspaces in the feature space, we utilize a divide-and-conquer approach for model-agnostic rule extraction. In the Model Compiler, we first propose a tree-based clustering algorithm to partition the feature space into subspaces, then design a decision boundary estimation mechanism to approximate the source model in each subspace. The Model Interpreter interprets predictions by important attributes to aid network operators in understanding the predictions. The Model Debugger conducts incremental updating to rectify errors by only fine-tuning rules on affected subspaces, thus reducing maintenance costs.
We implement a prototype using physical hardware, and experiments demonstrate its superior performance of 100 Gbps throughput, great interpretability, and trivial updating overhead.
\end{abstract}


\section{Introduction}
The network intrusion detection system (NIDS) has been a crucial network security infrastructure for decades. Among its various categories, anomaly-based NIDS (A-NIDS) that works in an unsupervised manner is drawing more attention. 
Compared to supervised methods, this type of methods is more promising because 1) it eliminates the need of attack data for training; 2) it does not rely on predefined threat models, improving the detection of unforeseen anomalies.
With the advances of various unsupervised machine learning (ML) and deep learning (DL) models, many A-NIDS approaches \cite{kitsune,unlearn,zerowall,iotargos,whisper,ardiot,iotensemble,hypervision} have been proposed. While these approaches have demonstrated considerable malicious traffic detection capability, they suffer from a few limitations that hinder their practical use.

\noindent
\textbf{Low throughput and high delay.} 
The inference speed of ML/DL models can hardly catch up with the soaring speed of today's high-throughput networks (e.g., 100 Gbps). As a consequence, most of these methods can only work in an off-path deployment fashion (e.g., deploy on a GPU server on control plane), causing prolonged response time.

\noindent
\textbf{Limited interpretability.} 
Due to the black-box nature of many ML/DL models, network operators are often reluctant to trust the high-stake decisions made by these models, whose output are typically scores/labels unintuitive for humans.

\noindent
\textbf{High updating overhead.}
Most methods require retraining to update models (e.g., to solve false positives), which is time-consuming and induces extra overhead.

Meanwhile, the recent advance in programmable data plane empowers in-network intelligence (INI), and opens up the possibility of fully deploying ML/DL-based NIDS on switching ASICs for line-speed processing. 
Prior works have realized the deployment of a series of learning models for intrusion detection, such as neural networks \cite{net2net,bnn_ids,jointNIDS}, SVM \cite{iisy,planter2}, decision trees \cite{iisy,IN_classification}, and ensemble models \cite{pforest,planter}.
However, existing INI solutions have two problems: 1) the vast majority are only applicable to specific models; 2) the only one general INI method (Mousika \cite{mousika}) conducts model-agnostic translation to decision trees for deployment, while it only supports supervised models. Currently, there is no work proposing general INI methods for unsupervised A-NIDS.


Designing a general INI framework of A-NIDS faces several challenges.
\textit{First}, most existing rule extraction methods (e.g., \cite{mousika,redt,trustee,blackbox_extract}) are supervised, 
mainly due to their heavy reliance on rule extraction models that inherently require labeled data for each class (e.g., CART decision trees).
\textit{Second}, existing methods often lack support for incremental updates, as their rules are obtained from the overall distribution of data, which needs to be re-estimated even for small updates.
Yet retraining models can lead to significant changes in the extracted rule set, requiring reinstallation of a large number of rules on the switch. 
\textit{Third}, A-NIDS typically requires more complex features, especially flow-level statistics (e.g., packet size mean/variance), to achieve sufficient accuracy without using labels. However, current switching ASICs usually have arithmetic constraints (e.g., do not support division) that restrict the support for complex features. As such, most prior methods (e.g., \cite{iisy,mousika,planter2}) only consider packet-level features.

In this paper, we present Genos, a general in-network deployment framework for various A-NIDS models. We adopt a systematic approach, referring to the tools of program development, and design three modules: Model Compiler, Model Interpreter, and Model Debugger.
The Model Compiler treats an A-NIDS as the ``source model'', converts it into a rule set as the ``intermediate representation'', and forms a set of P4 tables as the ``object representation'', enabling the efficient deployment of A-NIDS within programmable switches' data plane.
The Model Interpreter comes into play when an anomaly is detected. It utilizes the decision logic of the extracted rules and provides explanations for the identified anomalies, which assists network operators in understanding the underlying reasons for the predictions.
To address false positives, the Model Debugger can identify the rules responsible for incorrect decisions and incrementally generate a limited number of new rules to rectify the errors.

Due to the multimodal nature of benign data (e.g., a server often supports multiple services), benign samples often locate in multiple disjoint subspaces within the feature space. Following this intuition, Genos resolves the aforementioned challenges with three key designs. 
First, we design a\textit{divide-and-conquer rule extraction} method. Specifically, we propose a\textit{ Score Clustering Tree}, which partitions the feature space according to scores generated by the unsupervised source model into subspaces, each containing samples of similar normality. Then, for each subspace, we design a \textit{Decision Boundary Estimation} approach to obtain axis-aligned rules that accurately approximate the decision boundaries of the source model. 
Second, we realize incremental updating to rectify errors, which locates errors to the granularity of subspaces, and directly fine-tunes the extracted rules only on the affected subspaces. In this way, we avoid retraining the source model and reduce the number of rules to be reinstalled.
Third, we realize a feature extractor on the data plane that supports the acquisition of bidirectional flow-level features, especially finding a workaround to handle the comparison of complex features that cannot be easily computed by switching ASICs.

We implement a prototype on a commodity programmable switch and evaluate the deployment of four distinct A-NIDS models. Compared to four prior works (including Mousika), our rule extraction method achieves better results of 97.79\% fidelity and 98.80\% detection accuracy.
In general, Genos can achieve about 100 Gbps throughput, interpretable detection results, 
and trivial model updating overhead. 

We summarize our contributions as follows:
\begin{itemize}[topsep=0pt,itemsep=0pt,parsep=0pt,partopsep=0pt]
\item A general in-network A-NIDS framework achieving better throughput, interpretability, and incremental update.
\item 
To the best of our knowledge, it is the first work to realize the model-agnostic rule extraction of A-NIDS models to P4 tables in a fully unsupervised manner.
\item A prototype on hardware, realizing flow-level feature extraction within switching ASICs' limited operations.
\end{itemize}

\section{Background and Motivation}
\subsection{Anomaly-based Network Intrusion Detection (A-NIDS)}
A-NIDS is a promising category of NIDS as it requires no attack data and meanwhile can better detect unseen attacks.
A-NIDS approaches are typically built upon unsupervised learning models. For example, Mirsky et al. \cite{kitsune} and Li et al. \cite{iotensemble} both use autoencoders for intrusion detection. Binbusayyis et al. propose an unsupervised NIDS combining convolutional autoencoder and one-class SVM \cite{ocsvm_ids}. In \cite{horuseye}, the authors propose an A-NIDS approach based on isolation forest. 

Formally, for a $d$-dimensional feature space $\mathcal{X}$, given a stationary distribution $\mathcal{D}$ of normal traffic, an A-NIDS can be abstracted as a function $f$ that estimates the probability density function of benign data distribution, i.e., $f(\boldsymbol{x}) \approx \mathrm{P}_{\mathcal{X} \sim \mathcal{D}}(\boldsymbol{x})$, and detects anomalies via a threshold $f(\boldsymbol{x}) < \varphi$. In practice, this probability can be translated to different criteria of \textit{anomaly scores}, such as reconstruction error of autoencoders and average traversing path length of isolation forests.

Though A-NIDS has many advantages, there are also some limitations that hinder their practical use:
1) low throughput: on control plane, even extremely efficient approaches like \cite{whisper} can only achieve 10 Gbps-level throughput; 2) limited interpretability: most approaches only output a score (e.g., mean squared error \cite{zerowall,ardiot}) or a label (e.g., one-hot label \cite{iotargos}) that cannot explain anomalies in terms of important attributes; 3) high overhead of updating: cumbersome retraining process is required to update models.

\subsection{P4 Switch and In-Network Intelligence}
P4 switches \cite{p4} are equipped with programmable switching ASICs. They allow customized processing logic inside match-action pipelines through P4 programs, while reaching extraordinary processing speed and throughput (e.g., Tbps). 
Such advantages prompt the research on deploying ML/DL models directly on P4 switches, which is referred to as \textit{in-network intelligence} (INI). 

To realize INI, several computation constraints of P4 need to be considered, such as lack of support for division, float operations and loop operations.
Recent works have realized the deployment of different ML models. For example, Xiong et al. transform four ML models into match-action tables to realize INI \cite{iisy}. 
Some other works focus on tree-based models for INI as they naturally fit the match-action logic \cite{IN_classification,pforest,planter,planter2}.
As the ML community flourishes, ML/DL models will continue to evolve, and the race of deploying new models will never end.
Instead of designing model-specific INI solutions, Mousika \cite{mousika} first proposes the concept of \textit{general INI}. It adopts the technique of knowledge distillation \cite{redt} to translate the knowledge of well-trained models into binary decision trees, enabling indirect deployment of complex models and better flexibility over model-specific methods. 
However, Mousika can only work for supervised models and cannot be applied to A-NIDS approaches that are unsupervised.

\subsection{Challenges of General INI for A-NIDS}
\begin{figure}[t]
    \centering
    \subfloat[Autoencoder]{
    \includegraphics[width=0.4\linewidth]{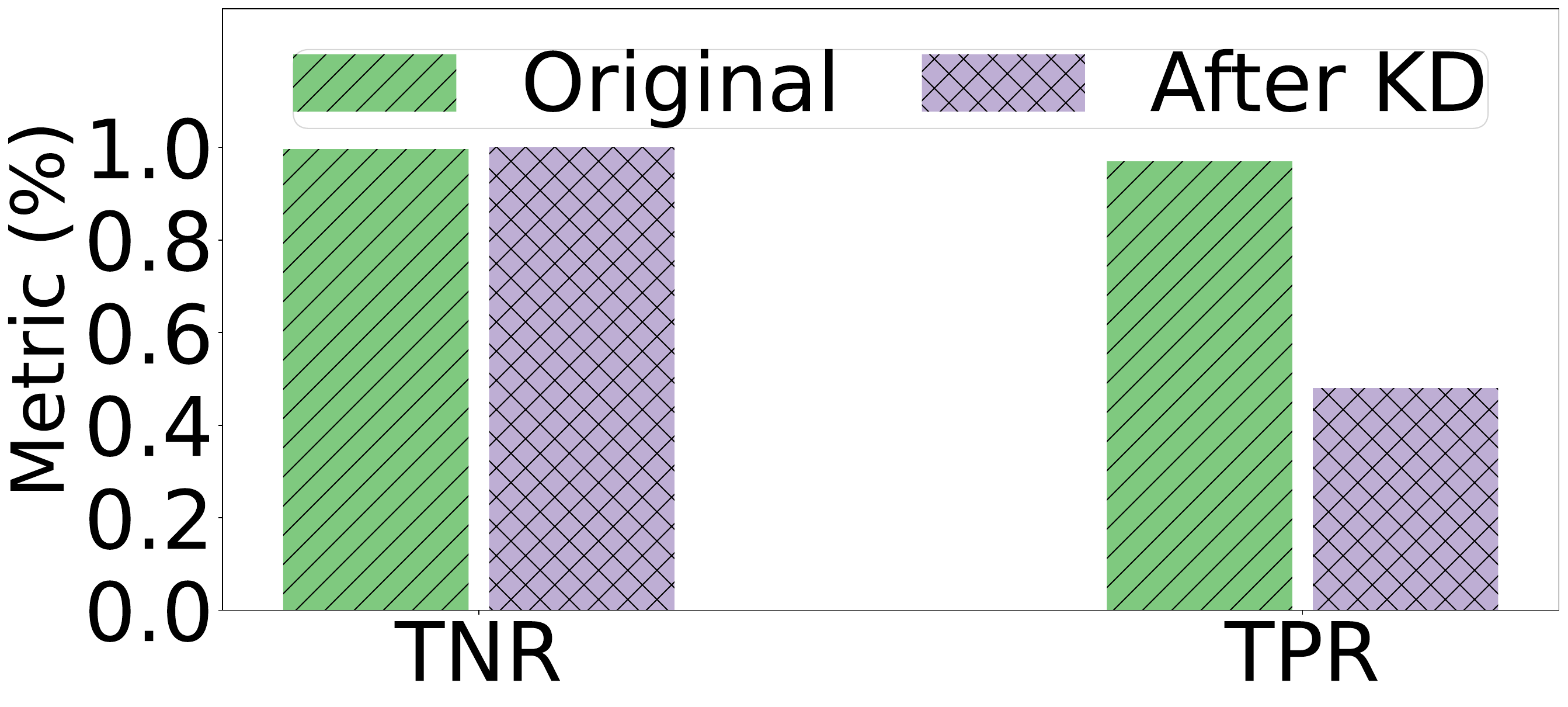}
    }
    \hfill
    \subfloat[Variational autoencoder]{
    \includegraphics[width=0.4\linewidth]{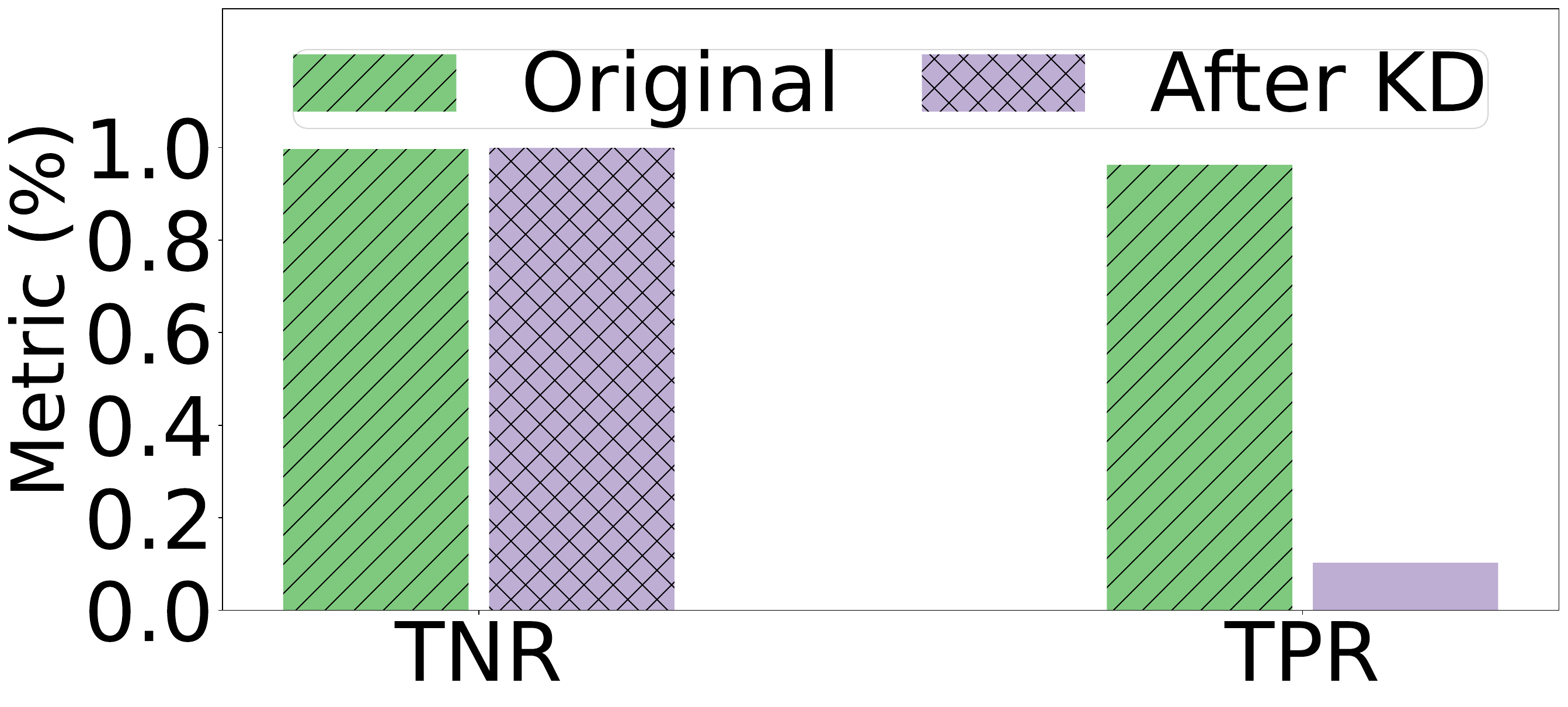}
    }
    \caption{
    Rules that extract A-NIDS using knowledge distillation (KD) and only benign data suffer huge accuracy loss.}
    \label{fig:mtv_ch1}
\end{figure}

To the best of our knowledge, research on general INI for unsupervised A-NIDS still remains blank. We attribute this vacancy to the following unresolved challenges:

\noindent
\textbf{Unsupervised rule extraction.}
The key technique used by Mousika, the general framework for supervised models, is knowledge distillation. In a larger scope, it belongs to model-agnostic \textit{rule extraction} aiming to translate black-box models into interpretable rules. 
Most of existing methods use CART decision trees as the translation target for rule extraction \cite{decision_list,blackbox_extract,trustee}. They
require labeled data for each class to precisely determine the decision logic of the models, which cannot meet A-NIDS’ requirement of only using unlabeled benign data. Yet forcing supervised rule extraction methods in a one-class scenario will cause severe accuracy loss.
For example, in Fig. \ref{fig:mtv_ch1}, compared to the original models, decision trees obtained by knowledge distillation show drastic performance degradation in detecting attacks (i.e., low true positive rate).

\setlength{\textfloatsep}{.75\baselineskip}
\setlength{\intextsep}{.75\baselineskip}

\begin{figure}[t]
    \centering
    \subfloat[Before update]{
    \includegraphics[width=0.4\linewidth]{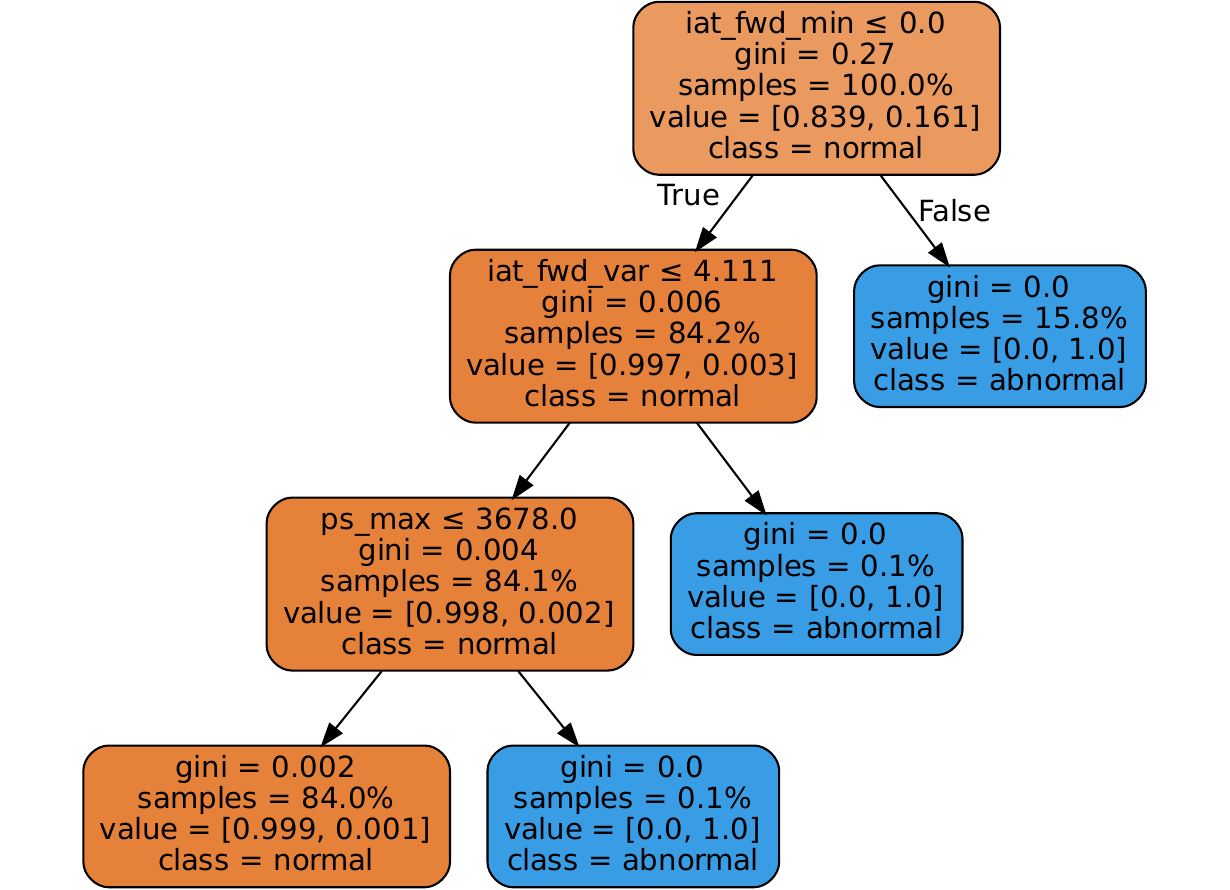}
    }
    \hfill
    \subfloat[After update]{
    \includegraphics[width=0.51\linewidth]{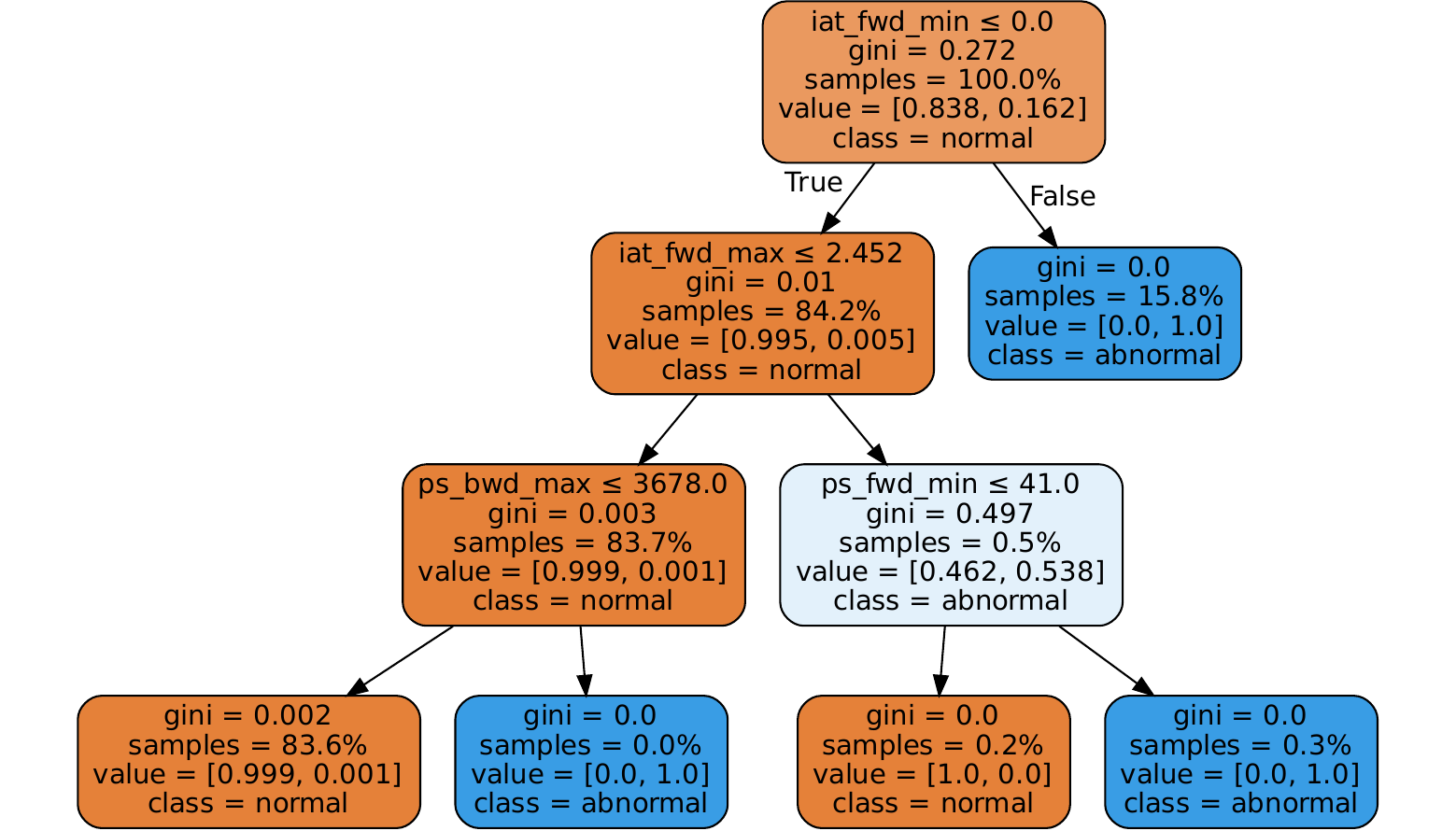}
    }
    \caption{Updating false positives with only 0.24\% of data causes changes in tree structure and splitting criteria.}
    \label{fig:mtv_ch3}
\end{figure}

\noindent
\textbf{High overhead of update.}
Deployed models sometimes need updates, e.g., when network operators find alarms are false positives and expect to reduce similar cases. 
Typically, we need to retrain the models and re-obtain the rules. However, many rule extraction methods \cite{redt,softdt,blackbox_extract,trustee} adopt decision trees, which inherently do not support incremental updates. Retraining tree models, even using a small scale of data, can cause remarkable changes in tree structure and rules. Fig. \ref{fig:mtv_ch3} illustrates such an example. As such, a large number of deployed rules have to be deleted and reinstalled to fix a small number of errors. Worse, this process can cause certain downtime of a switch, increasing the risk of being infiltrated.

\noindent
\textbf{Acquisition of flow-level features.} 
A-NIDS approaches usually need sophisticated features to precisely represent normal traffic, particularly flow-level features. Most methods (e.g., Planter \cite{planter}, Mousika \cite{mousika}) only support packet-level feature extraction on P4 switches. A more recent work \cite{netbeacon} first realizes stateful flow-level features, including statistics like mean and variance, by using bit shift operations to replace division operations. However, this approach can only summarize a flow at a fixed inference length of a power of 2, which can be insufficient for detecting time-related attacks. For example, there might be little difference between the bytes of the first four packets of a normal HTTP request and a CC attack.

\section{Overview}
This paper proposes a novel framework, Genos, which implements a complete life cycle for general in-network deployment of A-NIDS.
We observe that benign samples often locate in multiple disjoint subspaces within the feature space, due to the multimodal nature of benign data. Therefore, we adopt a divide-and-conquer approach, and extract rules on the granularity of subspace. This not only enables accurate approximation of the source model, but also allows incremental updating by only fine-tuning rules on the affected subspaces.

\begin{figure}[t]
    \centering
    \includegraphics[width=\linewidth]{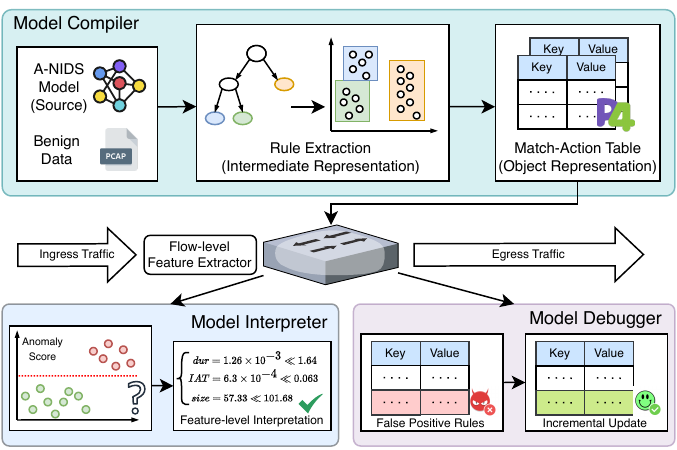}
    \caption{Overview of Genos.}
    \label{fig:overview}
\end{figure}

Fig. \ref{fig:overview} shows the overview.
We refer to the concepts in program development and design three modules running on control plane:
1) Model Compiler that translates an A-NIDS model into P4 tables by rule extraction;
2) Model Interpreter that explains important attributes for decisions; 
3) Model Debugger that fixes wrong decisions by incremental updates.

\noindent
\textbf{Model Compiler.}
This module achieves the translation of a black-box A-NIDS model into a set of rules, which are subsequently transformed into match-action tables for efficient in-network deployment.
We propose a model-agnostic rule extraction algorithm that addresses the challenge of unsupervised model extraction. 
The algorithm comprises a Score Clustering Tree to partition the feature space into subspaces, and a Decision Boundary Estimation approach to determine the decision rules of the source model on each subspace.

\noindent
\textbf{Model Interpreter.}
This module interprets the opaque output of A-NIDS models (e.g., anomaly scores) by feature importance. Considering that rule extraction inherently provides a global explanation for the decision logic of the source model, we realize an effective and efficient local explanation method, i.e., explaining the detection of one anomaly at a time, based on the feature deviations of extracted rules.

\noindent
\textbf{Model Debugger.}
Thanks to our divide-and-conquer rule extraction, we design this module that can pinpoint the rules producing errors and incrementally update them.
We develop an \textit{excluding} mode for the scenarios where data go into some subspace misidentified as anomalies inherently by the source model, and a \textit{patching} mode for the scenarios where errors are attributed to insufficient generalization of the rule in a certain subspace. As a result, only a small number of rules from the affected subspaces need to be reinstalled on switches to increase accuracy, reducing the overhead of updates.

Besides, Genos realizes bidirectional flow-level feature extraction on the data plane. We manage to compare mean and variance values to rule thresholds without division. Particularly, we design an adaptive timeout mechanism for flow length determination, executing \textit{active} timeout for long lived flows and \textit{inactive} timeout for burst flows, promoting more timely and accurate representations of flows.


\section{Framework Design}
\subsection{Model Compiler}

\subsubsection{Design Goal}
When designing the Model Compiler, we aim to derive an in-distribution rule set $\mathcal{C} = \{C_1, C_2, ...\}$ from a trained A-NIDS model $f$, utilizing its anomaly threshold $\varphi$ and benign training data $\boldsymbol{X}$. To ease the final deployment on P4 switches, we focus on axis-aligned rules that involve straightforward comparison and exclude rules of other formats to avoid additional calculations (e.g., linear models \cite{lime}). Each rule $C = ... \wedge (x_i \bot \upsilon_i) \wedge ... \wedge (x_j \bot \upsilon_j)$ represents a conjunction of feature constraints, where $\upsilon_i$ denotes the bound for the $i$-th dimension; $\bot \in \{ \leq, > \}$ denotes the comparison. We represent a data sample satisfying a rule as $\boldsymbol{x} \in C$. 
The anticipated outcome entails the derived rules to exhibit a substantial fidelity to the source model, implying a comparable coverage of benign data and a similar detection rate of anomalies:
\begin{equation}
    \arg\min_{\mathcal{C}} \mathcal{L}_{\mathcal{X} \sim \mathcal{D}}(\mathcal{C}, f, \varphi) + \mathcal{L}_{\mathcal{X} \nsim \mathcal{D}}(\mathcal{C}, f, \varphi),
\label{eq:goal}
\end{equation}
where $\mathcal{D}$ is the stationary distribution of benign data, and $\mathcal{L}$ is a loss function to measure fidelity in a certain space.

\subsubsection{Rule Extraction}
The biggest challenge in solving the problem in \eqref{eq:goal} is the minimization of the second term. Without labeled abnormal samples, this term is neither deterministic nor can be easily estimated by sampling points in a high-dimensional large space.
The essence of the challenge comes from the multimodal nature of normal data. 
For example, a server supports multiple services such as web, email, and database, each represented by distinct features that are possibly located in separate regions within the feature space. The limited transition between these regions makes axis-aligned rules incompetent in accurately approximating the decision boundary of the source model.

Motivated by the above intuition, we design a divide-and-conquer strategy.
The primary concept revolves around partitioning the space into subspaces that encapsulate normal data with more compact distributions and then using axis-aligned rules to approximate the source model on each subspace.
To this end, we design a tree model for problem breakdown and an estimation approach for problem resolution. 

\textbf{Score Clustering Tree.}
We notice that, even in the absence of labeled data, the output of A-NIDS models (i.e., anomaly scores) can serve as a valuable indicator to guide the partitioning of the feature space into subspaces for compact distribution of normal data. 
Built upon the CART decision tree \cite{cart}, Score Clustering Tree (SCT) introduces a new splitting criterion.
Given the data $\boldsymbol{N}$ at a tree node, SCT first obtains the output of the source model $f(\boldsymbol{x})$ for each $\boldsymbol{x} \in \boldsymbol{N}$. SCT seeks a splitting point $s$ for the node that maximizes the gain:
\begin{equation}
    \arg\max_{s} I(\boldsymbol{N}) - \frac{|\boldsymbol{N}_l|}{|\boldsymbol{N}|} I(\boldsymbol{N}_l) - \frac{|\boldsymbol{N}_r|}{|\boldsymbol{N}|} I(\boldsymbol{N}_r),
\end{equation}
where $s$ includes the splitting feature and threshold, $\boldsymbol{N}_l$ and $\boldsymbol{N}_r$ are the data split to the left and right child nodes, respectively, and $|\boldsymbol{N}|$ denotes the number of data samples. $I$ is the Gini index calculated by $1 - \sum p_j^2$, where $p_j$ originally is the probability of each class on this node. Given that we only have unlabeled benign data, we modify $p$ to be the average output scores of the source model:
\begin{equation}
    p = \frac{1}{|\boldsymbol{N}|} \sum f(\boldsymbol{x}), \forall \boldsymbol{x} \in \boldsymbol{N}.
\end{equation}
Unlike clustering techniques that rely solely on Euclidean distance (e.g., K-means), our approach takes the predictions of the source model to acquire a more profound understanding of benign data distribution, thus can more effectively categorize related benign data into one subspace. A tree splits nodes until it satisfies one of the conditions: i) the number of data samples at the node $|\boldsymbol{N}| = 1$; ii) output scores between any of the data samples at the node are below a limit, i.e., $\forall \boldsymbol{x}^{(i)}, \boldsymbol{x}^{(j)} \in \boldsymbol{N}$, $|f(\boldsymbol{x}^{(i)}) - f(\boldsymbol{x}^{(j)})| < \epsilon$; iii) it reaches the maximum depth $\tau$.

While our tree primarily serves as a space-splitting mechanism, it can also directly identify certain subspaces as anomalies. 
We define two types of leaf nodes in our tree: anomalous leaf nodes and unlabeled leaf nodes. The former is assigned only when all the data samples at a leaf node possess output scores lower than the threshold set by the source model, i.e.,
\begin{equation}
   y_{\boldsymbol{N}} = 1 \text{ iff } |\{\boldsymbol{x \in \boldsymbol{N}}; f(\boldsymbol{x}) < \varphi\}| = |\boldsymbol{N}|.
\end{equation}
With benign training data, such cases may arise when outliers are present within the data or if the source model exhibits false positives due to inadequate generalization. For the latter scenario, we will discuss how to rectify errors in the Model Debugger. For unlabeled leaf nodes, we use the following method to further extract their rules of normality.

\textbf{Decision Boundary Estimation.}
For each of the subspaces determined by a leaf node, we design an approach using axis-aligned rules to approximate the decision boundary of the source model. 
Let $\boldsymbol{X}_k$ represent the training data falling into the $k$-th leaf node. To begin with, we utilize the minimal hypercube $H_k$ as a reference to bound each dimension of the data samples in $\boldsymbol{X}_k$, which are identified as normal by the source model, thus ensuring $\mathcal{L}_{\boldsymbol{x} \in \boldsymbol{X}_k}(H_k, f, \varphi) = 0$. The minimal hypercube $H_k$ is encompassed by $2\times d$ axis-aligned hyperplanes, and can be described using the following rule:
\begin{equation}
\begin{aligned}
    H_k &= (\upsilon_1^- \leq x_1 \leq \upsilon_1^+) \wedge ... \wedge (\upsilon_d^- \leq x_d \leq \upsilon_d^+),\\
    \upsilon_i^- &= \min (x_i | f(\boldsymbol{x}) > \varphi, \boldsymbol{x} \in \boldsymbol{X}_k ),\\
    \upsilon_i^+ &= \max (x_i | f(\boldsymbol{x}) > \varphi, \boldsymbol{x} \in \boldsymbol{X}_k ),
\end{aligned}
\end{equation}
where $x_i$ denotes the $i$-th dimension of features in $\boldsymbol{x}$.

To explore the decision boundary, for the $i$-th dimension, we first conduct uniform sampling of $N_e$ data points on each hyperplane of the hypercube, denoted as $\boldsymbol{e}^{(1)}, \ldots, \boldsymbol{e}^{(N_e)} \in H_k \wedge (x_i = \upsilon_i), \upsilon_i \in \{\upsilon_i^-, \upsilon_i^+\}$. These data points are referred to as \textit{initial explorers}.
For each initial explorer $\boldsymbol{e}$, we further generate $N_s$ \textit{auxiliary explorers} in its vicinity, drawn from a truncated multivariate Gaussian distribution denoted as $\mathcal{N}(\boldsymbol{e}, \boldsymbol{\Sigma}, i)$. The center of sampling is an initial explorer $\boldsymbol{e}$, and the sampling radius is determined by the covariance matrix:
\begin{equation}
    \boldsymbol{\Sigma} = \text{diag}(\rho|\upsilon_1^+ - \upsilon_1^-|, \ldots, \rho|\upsilon_d^+ - \upsilon_d^-|),
\end{equation}
where $\rho$ is a hyperparameter to control the sampling radius. The sampling along the $i$-th dimension is half-truncated to ensure that only samples outside the hypercube are retained, in an attempt to extend the boundary.
With $N_e \times N_s$ auxiliary explorers in total, we query the source model and utilize Beam Search to select $N_e$ samples with the lowest output scores, indicating their proximity to the model boundary. 
To guarantee fast convergence towards the boundary, we refer to Fast Gradient Sign Method (FGSM) \cite{fsgm} for adversarial attacks and design an approximation approach for black-box scenarios. Its basic idea is to move towards the opposite direction to model training.
As an initial explorer $\boldsymbol{e}$ and its auxiliary explorer $\hat{\boldsymbol{e}}$ are spatially close, we can assume the model score is monotonous between the two points, and approximate the score's gradient at their midpoint by calculating the slope between the two points.
Similar to FGSM, we subtract the sign of gradient from the midpoint as the new initial explorer for the next iteration:
\begin{equation}
\begin{aligned}    
    &\boldsymbol{e}_{next} = \boldsymbol{e}_{mid} - \eta \cdot sign(\nabla_{\boldsymbol{e}_{mid}}), \\
    &\boldsymbol{e}_{mid} = \frac{\boldsymbol{e} + \hat{\boldsymbol{e}}}{2}, \nabla_{\boldsymbol{e}_{mid}} = \frac{\nabla f(\boldsymbol{\boldsymbol{e}_{mid}})}{\nabla \boldsymbol{\boldsymbol{e}_{mid}}} \approx \frac{f(\boldsymbol{e}) - f(\hat{\boldsymbol{e}})}{\boldsymbol{e} - \hat{\boldsymbol{e}}},
\end{aligned}
\end{equation}
where $sign(\cdot)$ is the sign function, and $\eta$ controls the stride. The iteration stops when it meets one of the two conditions:

\noindent
i) an auxiliary explorer $\hat{\boldsymbol{e}}_{last}$ that satisfies $f(\hat{\boldsymbol{e}}_{last}) < \varphi$ is found. 
For the $i$-th dimension, we establish a constraint to extend the boundary of the hypercube, i.e., $c_i = (x_i \leq \hat{e}_{last, i})$ if $\hat{e}_{last, i} > \upsilon_i^+$, or $c_i = (x_i > \hat{e}_{last, i})$ if $\hat{e}_{last, i} < \upsilon_i^-$;

\noindent
ii) it reaches the maximum iterations, suggesting the difficulty in moving towards the decision boundary by perturbing a particular feature dimension. We calculate the difference between the model output for the last auxiliary explorer $\hat{\boldsymbol{e}}_{last}$ and that of the first initial explorers $\hat{\boldsymbol{e}}$. If $|f(\hat{\boldsymbol{e}}) - f(\hat{\boldsymbol{e}}_{last})| < \delta$, we conclude that this dimension represents a \textit{contour line} for the source model. In this case, we do not produce any constraints for this dimension. Otherwise, we generate constraints in the same manner as those produced under the first condition. 

The final rule is obtained by taking the disjunction of the hypercube and the constraints on each dimension:
\begin{equation}
    C_k = H_k \vee (c_1 \wedge c_2 \wedge ... \wedge c_d).
\end{equation}

\setlength{\textfloatsep}{8pt}
\begin{algorithm}[t]
\SetAlgoLined
\SetKw{KwForIn}{in}
\SetKw{KwAnd}{and}
\SetKw{KwOr}{or}
\SetKw{KwNot}{not}
\SetKw{KwContinue}{continue}

\KwIn{A-NIDS model $f$ and threshold $\varphi$; dataset $\boldsymbol{X}$}
\KwOut{axis-aligned rule set $\mathcal{C}$}
Initialize an empty set $\mathcal{C}$\;
$T \gets \text{ClusteringTree}(\boldsymbol{X}, f, \varphi)$\;
\For{\text{leaf node} $\boldsymbol{N}$ \KwForIn $T$} 
{
    $C_{\boldsymbol{N}}, y_{\boldsymbol{N}} \gets \text{Root2LeafRule}(T, \boldsymbol{N})$\;
    \uIf{$y_{\boldsymbol{N}} = 1$}{
        $\mathcal{C}.\text{append}(\neg C_{\boldsymbol{N}})$\;
    }
    \uElse{
        $C_{be} \gets \text{BoundEstimate}(\boldsymbol{N}, f, \varphi)$\;
        $C \gets C_{\boldsymbol{N}} \wedge C_{be}$\;
        $\mathcal{C}.\text{append}(C)$\;
    }
}
\KwRet $\mathcal{C}$\;

\caption{Rule Extraction from A-NIDS}
\label{algo:rule_extract}
\end{algorithm}

The rule extraction algorithm is presented in Algorithm \ref{algo:rule_extract}. The result rule set comprises both the complementary rule for anomalous leaf nodes (lines 4$\sim$6), and the conjunction of the top-down tree rules and the rules obtained through boundary estimation for each unlabeled leaf node (lines 8$\sim$10).

\subsubsection{Translation to P4 Table}
\label{subsubsec:p4_table}
Due to the axis-aligned nature, our extracted rules can be readily translated into P4 tables using range matching for each feature dimension, as shown in Listing \ref{list:p4_table}. 
The majority of the table serves as an allowlist except for the rules derived from anomalous leaf nodes. Flows failing to match any rules are set as anomalies. 
We also adopt prior efforts \cite{mousika,netbeacon} to further encode range matching into ternary matching to enhance compatibility across switches. 

\begin{lstlisting}[caption=P4 table for extracted rules., label=list:p4_table, language=p4, mathescape]
table anids_rules {
    key = {
        meta.f_1: range;
        meta.f_2: range;
        ...
        meta.f_d: range;
    }
    actions = {set_benign, set_anomalous}
    default_action = {set_anomalous;}}
\end{lstlisting}

\subsection{Model Interpreter}
This module demystifies the opaque outputs of A-NIDS to improve human understanding and trust in model decisions. 
This process is known as \textit{local interpretation}, typically outputting the most important features related to a decision.
Existing works in this field \cite{local1,local2,local3_sec,local4_sec} often involve fitting a local linear model (e.g., lasso) to uncover the feature weights, which can be slow and unsuitable for delay-sensitive detection tasks, e.g., real-time online security analysis.
Our method can provide accurate and fast interpretations thanks to the inherent global interpretation provided by our rule extraction, which captures the decision logic of the source model and allows for the interpretation of individual decisions.

Considering a data sample $\boldsymbol{x}$, its prediction $y_{\boldsymbol{x}}$ and the extracted rule set $\mathcal{C}$, the goal of Module Interpreter is to output a feature importance vector $\boldsymbol{p}_{\boldsymbol{x}}$ to explain the prediction. The process is described in Algorithm \ref{algo:interpreter}. In line 2, we first pinpoint the rule corresponding to $\boldsymbol{x}$. As SCT splits the feature space into subspaces by leaf nodes, the rules obtained on each leaf node are disjoint, resulting in at most one rule matching a given data sample. 
For a normal prediction, its important feature should be a distinct space that anomalies are unlikely to fall in.
Thus, we use the reciprocal of each feature constraint range as the importance (line 5). For an abnormal prediction, as our rules describe the space of normality, the features that do not conform to the rules reveal the anomaly, and higher deviations indicate more anomalous. So we use the distance between the feature and its violated bound as the weight (line 7). If there is no constraint on a feature, we will not assign weights to this feature. All feature values and rule bounds are normalized so that the computed weights are comparable.

\begin{algorithm}[t]
\SetAlgoLined
\SetKw{KwForIn}{in}
\SetKw{KwAnd}{and}
\SetKw{KwOr}{or}
\SetKw{KwNot}{not}
\SetKw{KwContinue}{continue}

\KwIn{data sample ($\boldsymbol{x}, y_{\boldsymbol{x}}$), extracted rule set $\mathcal{C}$}
\KwOut{feature importance vector $\boldsymbol{p}_{\boldsymbol{x}}$}
Initialize a zero vector $\boldsymbol{p}_{\boldsymbol{x}}$\;
$C \gets \text{FindRule}(\boldsymbol{x}, \mathcal{C})$\;
\For{i \KwForIn $\boldsymbol{x}.size \text{ and } (\upsilon_i^- \leq x_i \leq \upsilon_i^+) \in C$}{
    \uIf{$y_{\boldsymbol{x}} = normal$}{
        $\boldsymbol{p}_{\boldsymbol{x}}[i] \gets 1 / (\upsilon_i^+ - \upsilon_i^-)$\;
    }
    \uElse{
        $\boldsymbol{p}_{\boldsymbol{x}}[i] \gets \text{ReLU}(x_i - \upsilon_i^+) + \text{ReLU}(\upsilon_i^- - x_i)$\;
    }
}
\KwRet $\boldsymbol{p}_{\boldsymbol{x}}$\;

\caption{Interpretation by extracted rules}
\label{algo:interpreter}
\end{algorithm}

\subsection{Model Debugger}
This module updates the deployed model to address errors, especially false positives. 
Unlike the conventional updating pipeline which involves model retraining and redeployment, our method circumvents these steps by directly operating on the extracted rules. This is made possible by our rule extraction algorithm, 
which isolates the model extraction from the overall feature space into small subspaces.

\begin{figure}[t]
    \centering
    \subfloat[Patching Mode on unlabeled leaf node]{
    \includegraphics[width=0.9\linewidth]{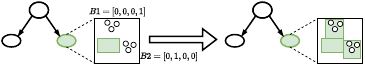}
    }
    \hfill
    \subfloat[Excluding Mode on anomalous leaf node]{
    \includegraphics[width=0.9\linewidth]{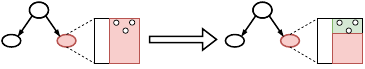}
    }
    \caption{Illustration of Model Debugger to fix false positives.}
    \label{fig:debugger}
\end{figure}

Suppose a network operator finds a batch of reported anomalies are false positives, denoted by $\boldsymbol{X}_{fp}$. 
Depending on the types of leaf nodes on which the samples are identified, our Model Debugger adopts two different modes (Fig. \ref{fig:debugger}):

\textbf{Patching Mode.}
False positives that happen on an unlabeled leaf node may result from our rule extraction not being sufficiently generalized to approximate the real decision boundary. Despite narrowing the overall feature space to small subspaces via our tree model, axis-aligned hyperplanes might still not perfectly fit the complex source model, especially when it is far from linear. In this case, our method incrementally generates a set of new rules, referred to as \textit{patches}.
False positive samples that arrive at the same leaf node may violate the rule of the leaf node with respect to different feature constraints. For each sample $\boldsymbol{x} \in \boldsymbol{X}_{fp}$, we use a $2\times d$-length bitmap $B_{\boldsymbol{x}}$ to record the dimensions that the sample does not conform to:
\begin{equation}
    B_{\boldsymbol{x}}[i] = 
    \begin{cases}
    01,& x_i > \upsilon_i^+, \\
    10,& x_i < \upsilon_i^-, \\
    \end{cases}
\end{equation}
where $\upsilon_i^-$ and $\upsilon_i^+$ represent the upper and lower bounds of the rule's constraint on the $i$-th dimension.

Next, we employ our Decision Boundary Estimation method to generate a new rule for samples that have identical bitmaps. These patch rules not only involve the false positives but also explore the boundary to a more reasonable space, improving the generalization of the rules and better mimicking complex and nonlinear decision boundaries of the source model.

\textbf{Excluding Mode.}
This mode applies when false positives occur on an anomalous leaf node due to the source model's misidentification of samples in this subspace. As a result, our tree model also labels this leaf node as anomalous. Since this subspace already falls outside the source model's decision boundary, our decision boundary estimation method cannot generate a patching rule.  
Instead, we use a minimal hypercube to exclude the area containing the false positive samples from the subspace, effectively creating a rule that removes the influence of these misidentified samples from the leaf node.

\subsection{Data Plane Implementation}
We implement the data plane of Genos on a Tofino programmable switch (1400 lines of P4 code). We detail the implementation of the flow-level feature extractor, which realizes the 30 flow-level features list in Table \ref{tab:fe} on the data plane.

\textbf{Bidirectional Flow Record.}
Like many control plane flow analyzers (e.g., \cite{cicflowmeter}), our implementation on the data plane also records bidirectional flows rather than unidirectional flows to better represent flow patterns. When a packet arrives, we parse its 5-tuple, packet size, and timestamp. Based on the packet direction (e.g., LAN to WAN), we use two hash tables to record the flow: a forward table and a backward table, where the keys are the forward and backward 5-tuples, respectively. 

\textbf{Stateful Storage.}
Each hash table entry points to a register array for stateful storage. Each register contains an incremental statistic of the flow. The first five statistics are space-related, including 1) packet counts $M$; 2) sum of packet sizes $LS_{s}$; 3) squared sum of packet sizes $SS_{s}$ (data plane supports the approximate square calculation); 4) maximum packet size $Max_{s}$; 5) minimum packet size $Min_{s}$.
In addition, we record the timestamp of the first packet $t_1$ and the timestamp of the last packet $t_{M}$ in the flow. Using these timestamps, we can calculate the flow duration and inter-arrival time between packets, which enable us to preserve four time-related incremental statistics: 1) sum of inter-arrival times $LS_{t}$; 2) squared sum of inter-arrival times $SS_{t}$; 3) maximum inter-arrival time $Max_{t}$; 4) minimum inter-arrival time $Min_{t}$.
The calculation of these time-related statistics can be implemented in two stages.

\textbf{Timeout Mechanism.}
Referring to the classic NetFlow \cite{netflow}, we design an adaptive timeout mechanism for flow length determination, which has two types of timeouts: 1) active timeout, which segments long flows to reduce persistent storage occupation, and is triggered when flow packet count reaches a threshold $m$; 2) inactive timeout, which identifies burst flows and is triggered when inter-arrival time exceeds a threshold value $\Delta$.
This mechanism offers dynamic flow length determination, enabling a more comprehensive understanding of flow patterns.

\textbf{Feature Acquisition.}
When a timeout occurs, we retrieve the statistics of the flow from the register array to build a feature vector for model inference. Since a register can only be accessed once in the pipeline, we utilize the \textit{resubmit} mechanism to obtain register values.
For features like L4 protocol, port, packet count, and maximum/minimum statistics, we can directly obtain them from the packet header and registers. 
The most challenging features are the mean and variance values, which cannot be computed by switching ASICs. 
We notice that their comparison to rule thresholds ($\upsilon_i$), i.e.,
\begin{equation}
\mu_i = \frac{LS}{M} \bot \upsilon_i, \;\;
    \sigma_i = \frac{SS}{M} - (\frac{LS}{M})^2 \bot \upsilon_i,
\end{equation}
where $\bot \in \{ \leq, > \}$, can be rearranged into:
\begin{equation}
   LS \bot M\cdot \upsilon_i,\;\; M\cdot SS - LS^2 \bot M^2\cdot \upsilon_i,
\end{equation}
where $M\cdot \upsilon_i$ and $M^2\cdot \upsilon_i$ can be computed in advance. In this way, we achieve the comparison without division operation.
We pre-encode $M\cdot \upsilon_i$ and $M^2\cdot \upsilon_i$ in P4 tables as rule thresholds for every possible value of $M$ (i.e., 1 to $m$ at most due to active timeout).
While this may increase the number of table entries, it remains acceptable as our extracted rules are efficient.

\begin{table}[t]
    \centering
    \scriptsize
    \renewcommand{\arraystretch}{.75}
    \caption{Flow-level features implemented on data plane.}
    \begin{tabular}{cccc}
    \toprule
    Attribute & Statistics & Direction & Number \\
    \midrule
    packet count & 1$\sim$$m$(active timeout) & fwd, bwd, both & 3 \\
    packet size & mean/max/min/var & fwd, bwd, both & 12 \\
    inter-arrival time & mean/max/min/var & fwd, bwd, both & 12 \\
    flow duration & microsecond & both & 1 \\
    destination port & 0$\sim$65535 & fwd & 1 \\
    L4 protocol & TCP or UDP & - & 1 \\
    \bottomrule
    \end{tabular}
    \label{tab:fe}
\end{table}

\section{Evaluation}
\subsection{Experimental Setup}
We use two benchmark datasets for network intrusion detection: CIC-IDS \cite{cicids} and TON-IoT \cite{ton-iot}. These datasets are in PCAP files, providing sufficient benign data (e.g., server traffic) and a wide range of realistic attack traffic (e.g., DDoS, scanning, botnet). The datasets are randomly split into training (40\%), validation (30\%), and testing (30\%) sets by flows.

We adopt four types of unsupervised models commonly as A-NIDS source models, including autoencoder (AE), variational autoencoder (VAE), one-class SVM (OCSVM), and Isolation Forest (iForest). These models are trained as well as hyperparameter calibration using only benign data.
The performance of these models is evaluated using the Area Under the ROC Curve (AUC) score.
Table \ref{tab:dataset} provides a description of the datasets, along with the AUC scores achieved by the A-NIDS models.
We use grid search to decide our hyperparameters.

\begin{figure*}[t]
    \centering
\includegraphics[width=0.3\linewidth]{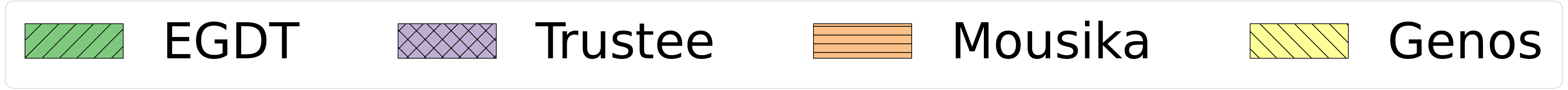}
\vspace{-.8cm}
\end{figure*}

\begin{figure*}[t]
    \centering
    \subfloat[Fidelity (CIC-IDS)]{
    \includegraphics[width=0.23\linewidth]{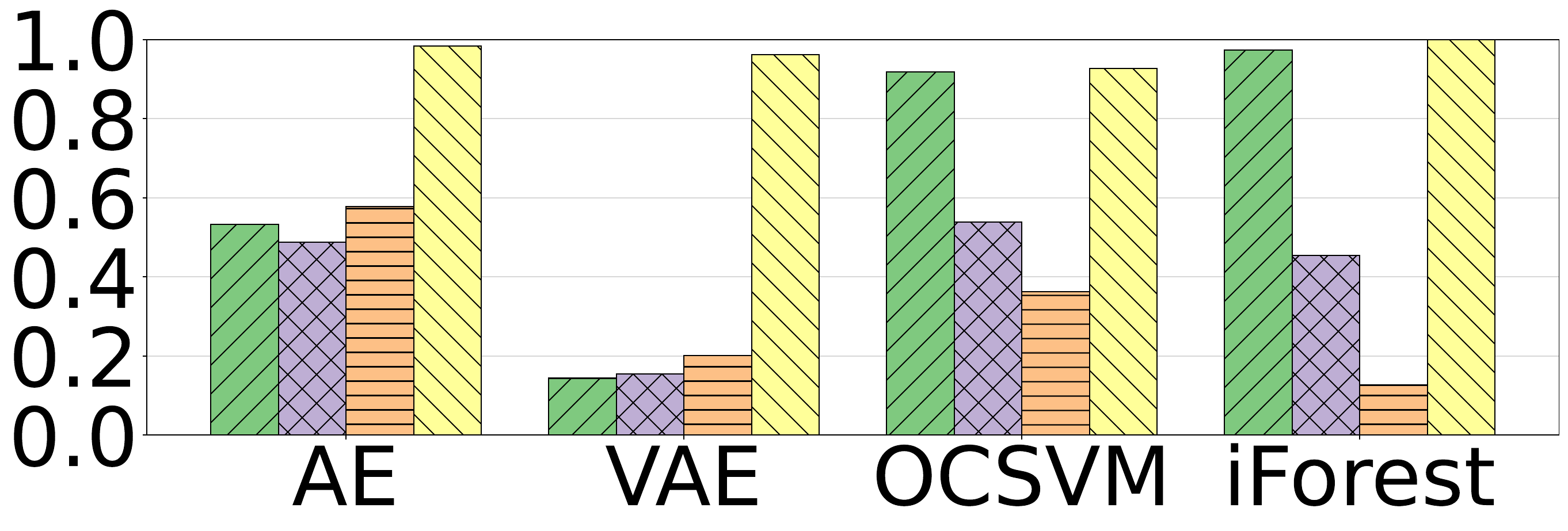}
    }
    \subfloat[Robustness (CIC-IDS)]{
    \includegraphics[width=0.23\linewidth]{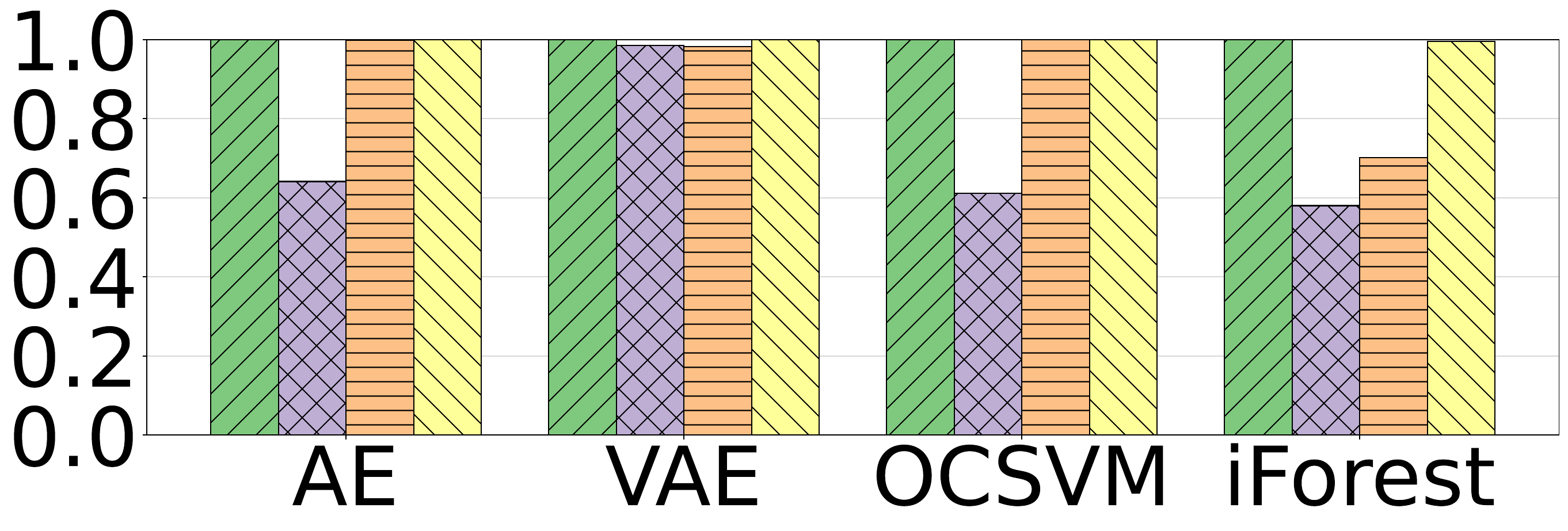}
    }
    \subfloat[TPR (CIC-IDS)]{
    \includegraphics[width=0.23\linewidth]{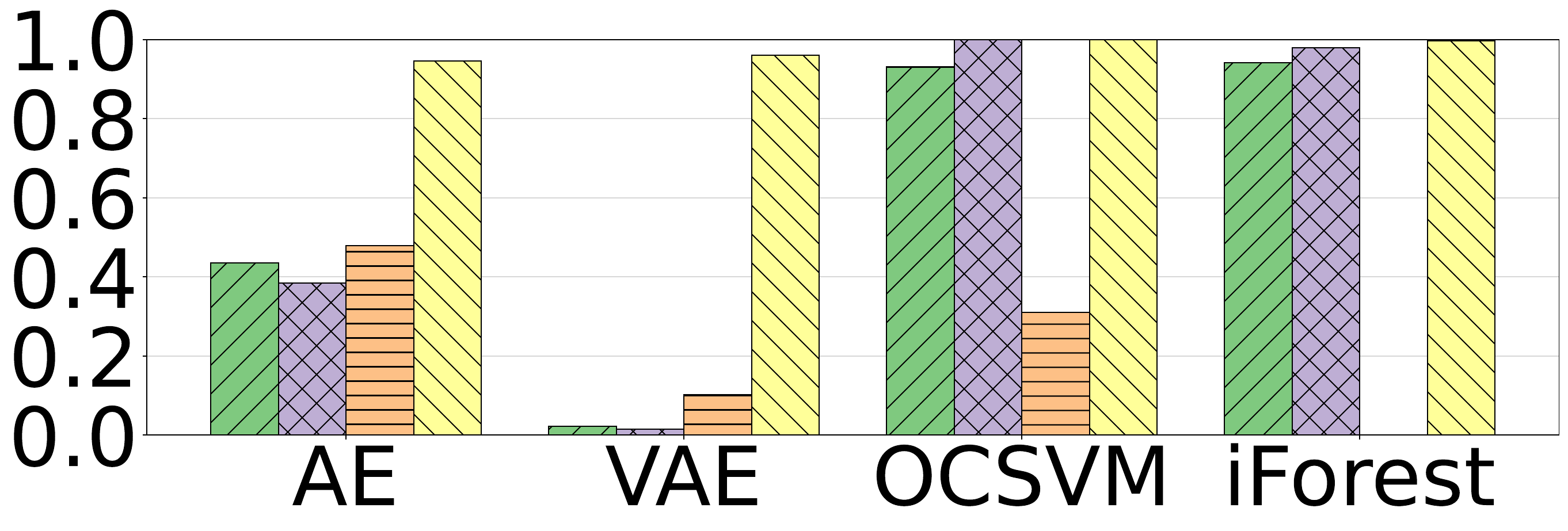}
    }
    \subfloat[TNR (CIC-IDS)]{
    \includegraphics[width=0.23\linewidth]{figure/results/CIC-IDS_TPR_nolegend.pdf}
    }
    \hfill
    \subfloat[Fidelity (TON-IoT)]{
    \includegraphics[width=0.23\linewidth]{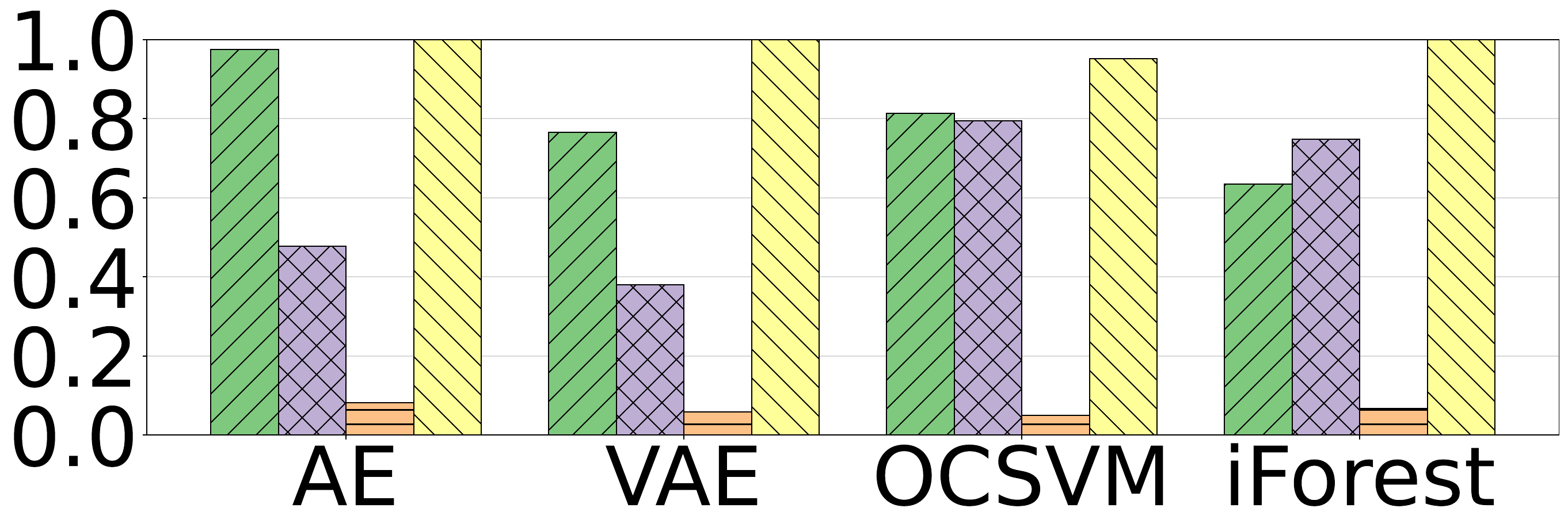}
    }
    \subfloat[Robustness (TON-IoT)]{
    \includegraphics[width=0.23\linewidth]{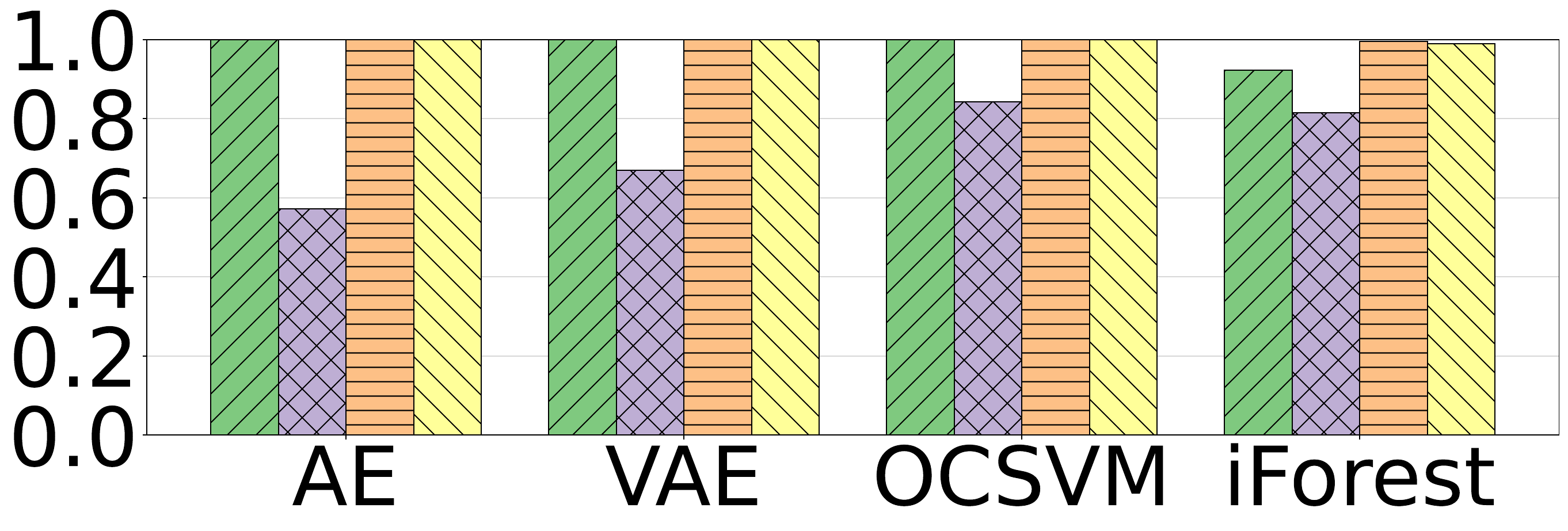}
    }
    \subfloat[TPR (TON-IoT)]{
    \includegraphics[width=0.23\linewidth]{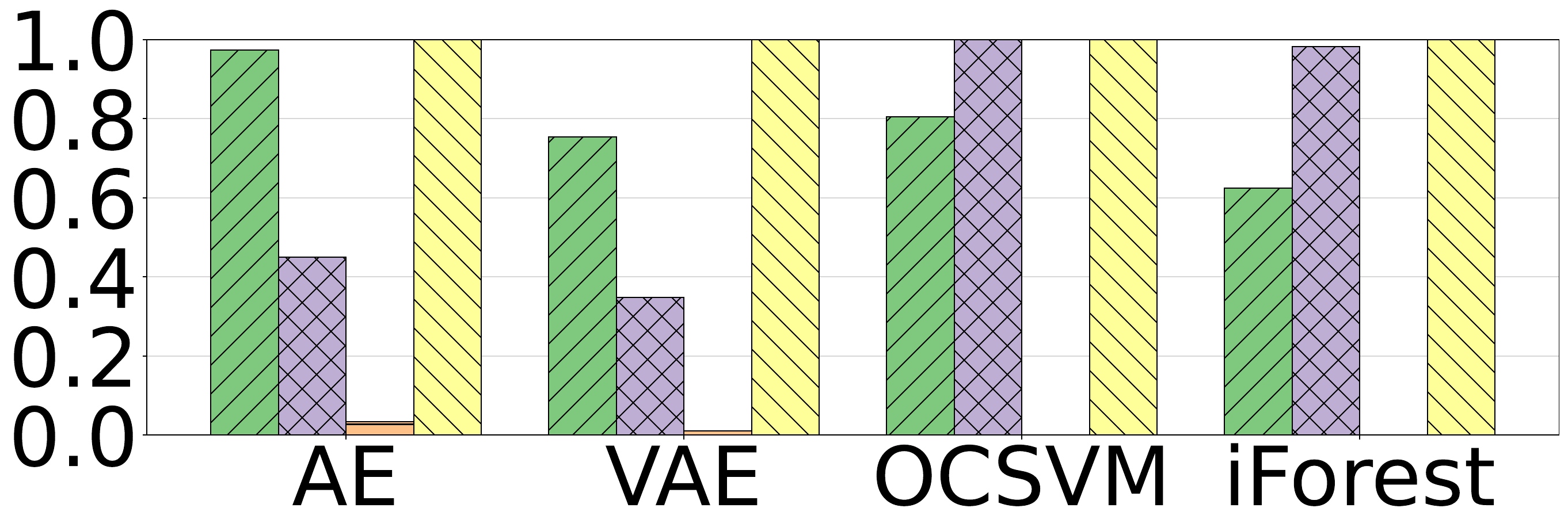}
    }
    \subfloat[TNR (TON-IoT)]{
    \includegraphics[width=0.23\linewidth]{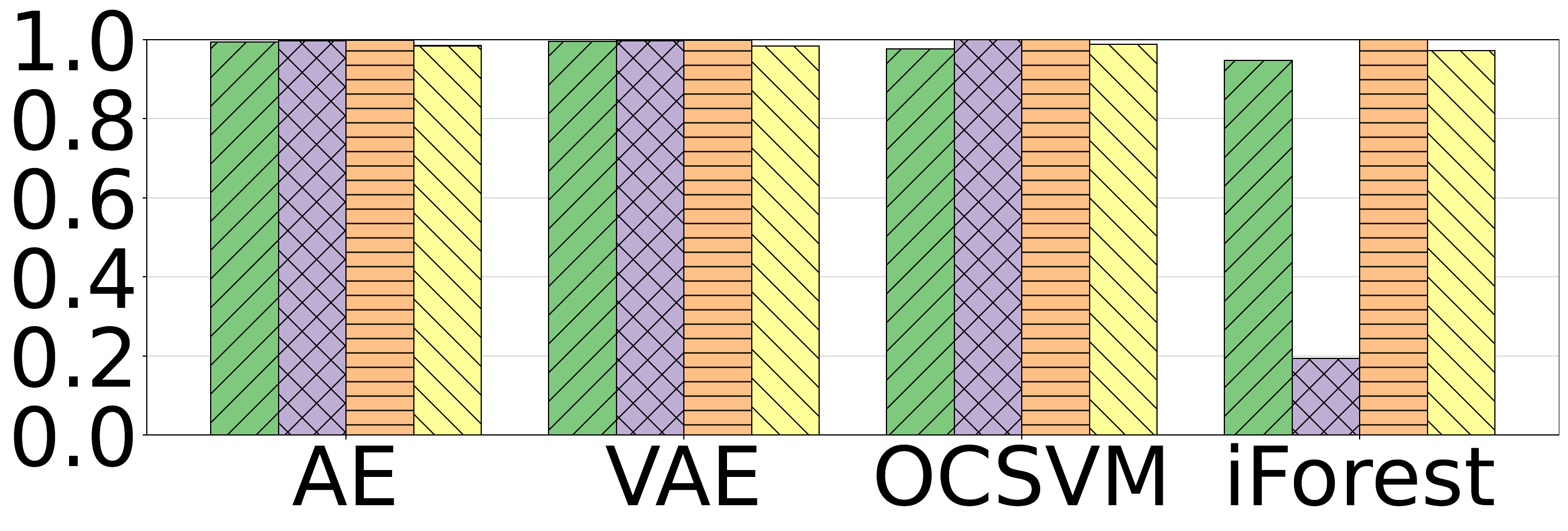}
    }   
    \caption{Comparison of rule extraction performance on four A-NIDS models using two traffic datasets.}
    \label{fig:rule_extraction1}
\end{figure*}

\begin{table}[t]
    \centering
    \scriptsize
    \tabcolsep=0.45cm
    \renewcommand{\arraystretch}{.75}
    \caption{Datasets and A-NIDS models.}
    \begin{tabular}{cccc}
    \toprule
    \multicolumn{2}{c}{Characteristics} & CIC-IDS & TON-IoT \\
    \midrule
    \multicolumn{2}{c}{PCAP size} & 40.1GB & 6.27GB \\
    \multicolumn{2}{c}{\#Attack types} & 6 & 9 \\
    \multicolumn{2}{c}{\#Normal samples} & 687,565 & 309,086 \\
    \multicolumn{2}{c}{\#Attack samples} & 288,404 & 893,006 \\
    \midrule
    \multirow{4}{*}{\makecell[c]{source model \\ AUC}} & AE & 0.9921 & 0.9998 \\
    ~ & VAE & 0.9901 & 0.9998 \\
    ~ & OCSVM & 0.9967 & 0.9993 \\
    ~ & iForest & 0.9879 & 0.9877 \\
    \bottomrule
    \end{tabular}
    \label{tab:dataset}
\end{table}

\begin{table}[t]
    \centering
    \scriptsize
    \tabcolsep=0.14cm
    \renewcommand{\arraystretch}{.75}
    \caption{Performance on each type of attacks.}
    \begin{tabular}{ccccccccc}
    \toprule
    \multirow{2}{*}{Attack} & \multicolumn{2}{c}{AE} & \multicolumn{2}{c}{VAE} & \multicolumn{2}{c}{OCSVM} & \multicolumn{2}{c}{iForest} \\
    ~ & TPR & TNR & TPR & TNR & TPR & TNR & TPR & TNR \\
    \midrule
    backdoor & 1.000 & 0.9772 & 1.000 & 0.9715 & 1.000 & 0.9857 & 1.000 & 0.9686 \\
    DDoS & 1.000 & 0.9832 & 1.000 & 0.9832 & 1.000 & 0.9851 & 1.000 & 0.9767 \\
    DoS & 1.000 & 0.9849 & 1.000 & 0.9849 & 1.000 & 0.9915 & 1.000 & 0.9831 \\
    injection & 1.000 & 0.9888 & 1.000 & 0.986 & 1.000 & 0.9795 & 1.000 & 0.9701 \\
    MITM & 1.000 & 0.9776 & 1.000 & 0.9669 & 1.000 & 0.9835 & 1.000 & 0.9582 \\
    brute force & 1.000 & 0.9816 & 1.000 & 0.9903 & 1.000 & 0.9874 & 1.000 & 0.9709 \\
    ransomware & 1.000 & 0.9962 & 1.000 & 0.9808 & 1.000 & 0.9789 & 1.000 & 0.9607 \\
    scanning & 1.000 & 0.9904 & 1.000 & 0.9865 & 1.000 & 0.9846 & 1.000 & 0.9605 \\
    XSS & 1.000 & 0.9862 & 1.000 & 0.9882 & 1.000 & 0.9774 & 1.000 & 0.9518 \\
    \midrule
    \#Rules & \multicolumn{2}{c}{29} & \multicolumn{2}{c}{27} & \multicolumn{2}{c}{17} & \multicolumn{2}{c}{17} \\
    \bottomrule
    \end{tabular}
    \label{tab:each_attack}
\end{table}

\subsection{Model-Agnostic Rule Extraction}
\textbf{Metrics.}
We use four metrics to evaluate the performance of rule extraction: 1) fidelity, the ratio of data samples on which the predictions of extracted rules are identical to the predictions of the source model; 
2) robustness, the ratio of data samples added with a perturbation, on which the predictions of extracted rules are identical to the predictions of the source model;
3) true positive rate (TPR) and 4) true negative rate (TNR), indicating accuracy in unbalanced data scenarios.

\textbf{Baselines.}
We compare our method to three baselines:

\noindent
1) \textit{Estimated Greedy Decision Tree (EGDT)} \cite{blackbox_extract}: It extracts a decision tree that actively samples new training points to mirror the computation performed by the source model. 

\noindent
2) \textit{Trustee} \cite{trustee}: It synthesizes high-fidelity and low-complexity tree models that can demystify black-box model decisions.

\noindent
3) \textit{Mousika} \cite{mousika}: It uses knowledge distillation to translate the knowledge of a complex model into a binary decision tree.

All baselines are comparable to ours since they are model-agnostic tree-based models that generate axis-aligned rules and can be deployed on programmable switches.
Extracting rules using these methods can only access benign datasets. 

Fig. \ref{fig:rule_extraction1} shows Genos achieves superior performance in all metrics.
First, it achieves the highest fidelity across all source models and datasets, with some scores even surpassing 0.999. It indicates that Genos can accurately capture the decision logic and precisely match the predictions of the source models.
Second, Genos achieves the highest TPR across all the source models and datasets, with a TPR of 1.00 for all the models on the TON-IoT dataset. This exceptional result signifies that our rules can accurately detect various anomalous traffic.
In contrast, most baselines fail to obtain adequate TPR since they rely on much anomalous data to determine decision boundaries, which may not be readily available in practice.
Third, the robustness of Genos is also commendable, with scores ranging from 0.9890 to 1.00, demonstrating its ability to handle perturbations in practical environments. 
Moreover, the TNR is consistently high (0.9715 to 1.00).


We also study the performance of our extracted rules in detecting each type of attacks. Table \ref{tab:each_attack} shows our rules can perfectly detect all types of attacks using any of the source models (TPR=1.000). The TNR also reaches a satisfying level (0.981 on average). 
Such a remarkable detection performance is owing to the high efficacy of our method, which extracts fewer than 30 rules, making it practical for deployment on resource-constrained network devices.


\subsection{Interpreting Decisions}
We compare the Model Interpreter of Genos to LIME \cite{lime}, a state-of-the-art model-agnostic explanation method. Following prior works (e.g., \cite{local3_sec}), we conduct a feature deduction test. Intuitively, if top-K features are selected as important for a decision, removing these features from this data sample would probably lead to misclassification by the source model. As we only access benign data for training, the selected features are essentially significant attributes of benign data. Thus, we use the Negative Flipping Rate (NFR) as the metric, which measures the ratio of the samples that are initially predicted as normal by the source model and are predicted as abnormal after nullifying the selected top-K features.

\begin{figure}
    \centering
    \subfloat[CIC-IDS]{
    \includegraphics[width=0.47\linewidth]{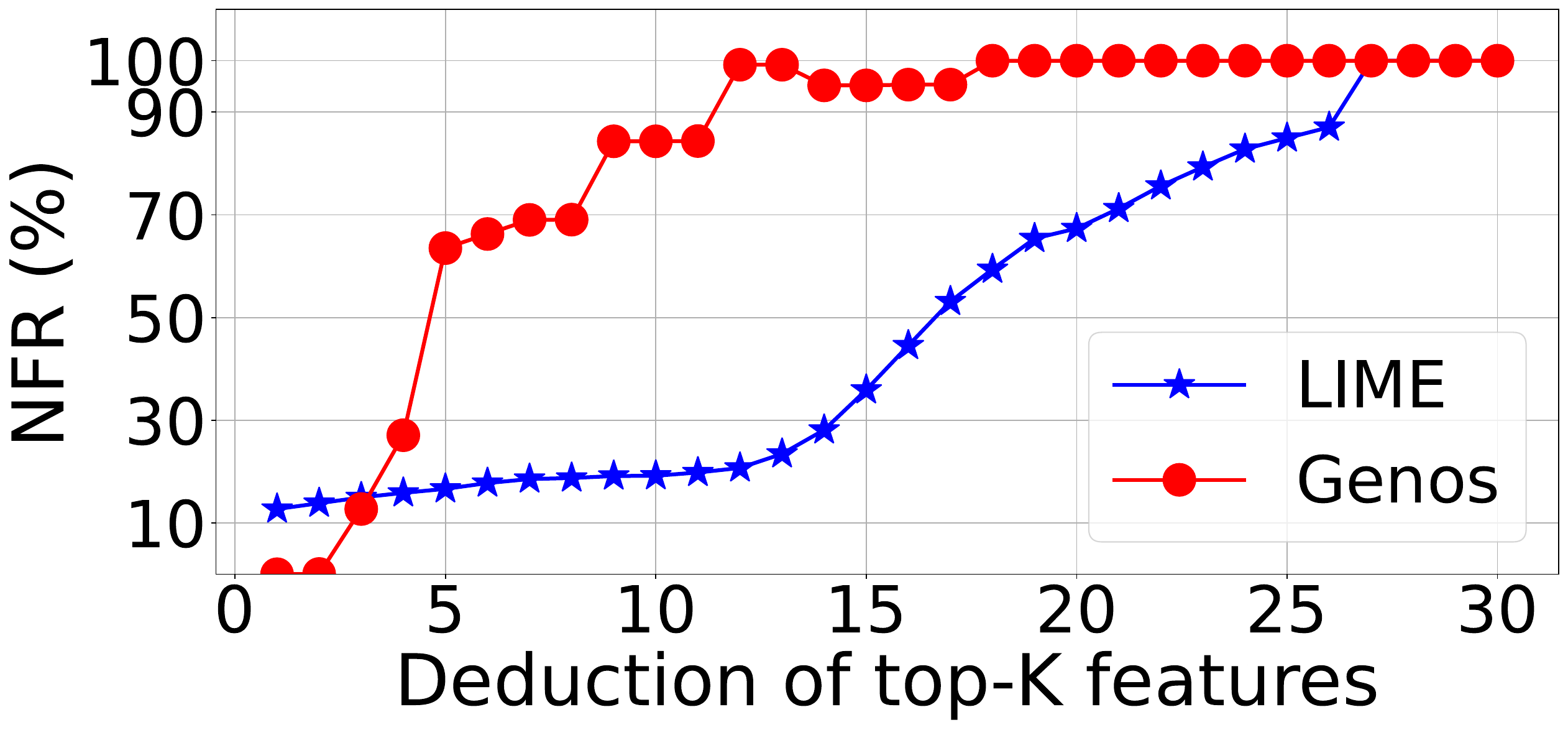}
    }
    \subfloat[TON-IoT]{
    \includegraphics[width=0.47\linewidth]{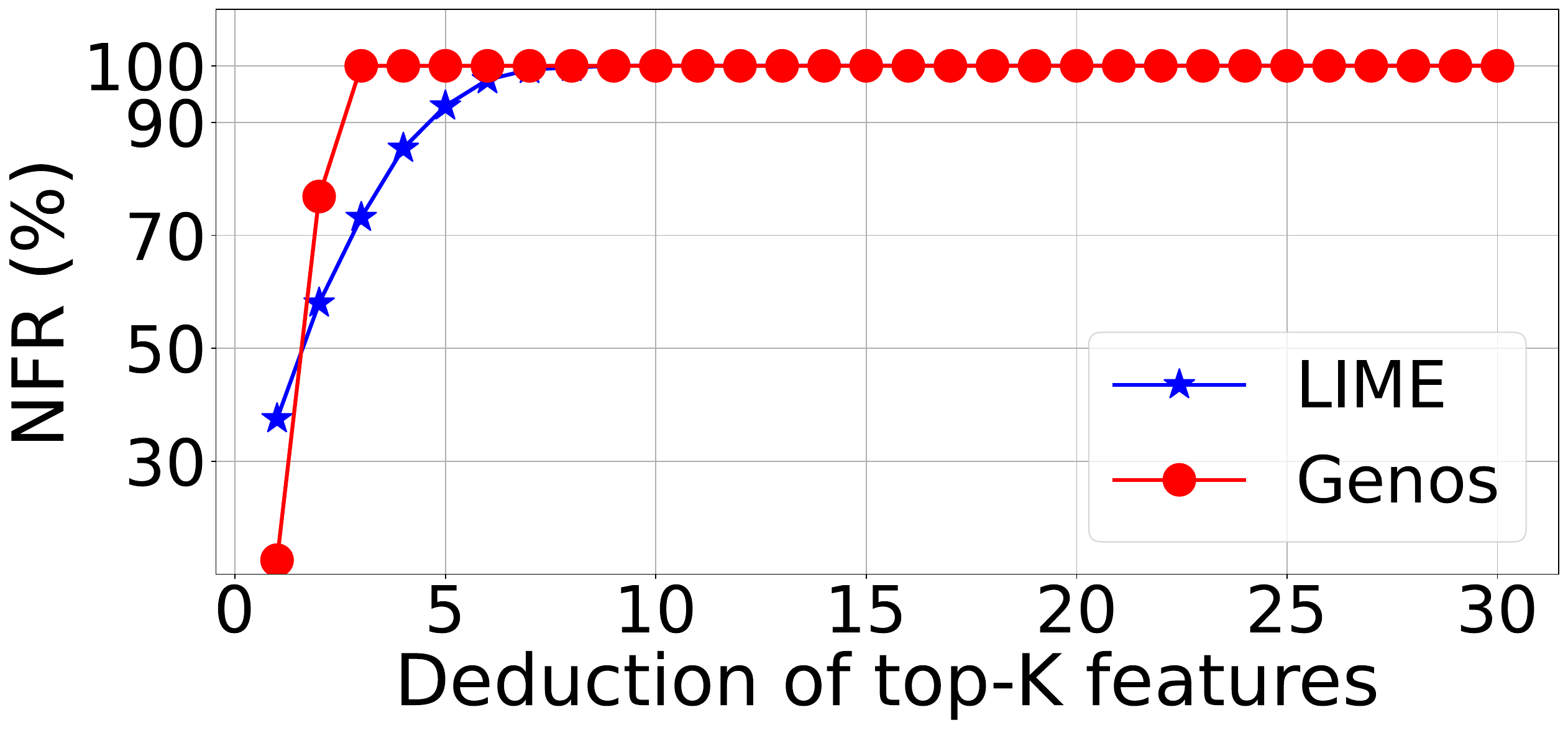}
    }    
    \caption{Feature deduction test on LIME and Genos.}
    \label{fig:feature_deduction}
\end{figure}

Fig. \ref{fig:feature_deduction} depicts Genos reaching high NFR faster than LIME on both datasets. On TON-IoT dataset, Genos attains 100\% NFR with only the top three features. 
In terms of efficiency, Genos is over 2000 times faster than LIME ($0.168\mu s$ v.s. $478.206\mu s$ on average) when interpreting a decision. 
We attribute the superiority to our rule extraction, which can well describe the feature ranges of normal data, and accurately and efficiently interpret decisions by deviations of feature values.

\begin{figure}[t]
    \centering
    \subfloat[]{
    \includegraphics[width=0.35\linewidth]{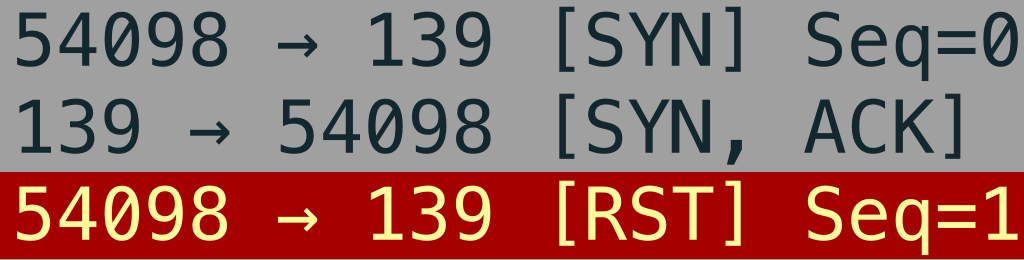}
    }
    \subfloat[]{
    \includegraphics[width=0.35\linewidth]{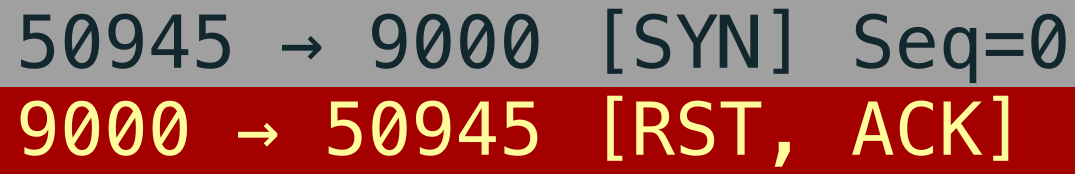}
    }
    \caption{Two scanning attack samples analyzed by Wireshark.}
    \label{fig:screenshot}
\end{figure}

Taking two samples from TON-IoT datasets that scan attack attempts as examples, Table \ref{tab:port_scan} presents the interpretation of Genos. One sample successfully scans an open port while the other fails.
As the normal traffic mainly consists of MQTT brokers with a limited range of ports, the wide scanning range of ports is a strong indicator for scanning attacks.
Another important feature is the flow packet count: one sample has three packets, while the other has two. Manual data analysis, as depicted in Fig. \ref{fig:screenshot}, also reveals a clear distinction: a successful scanning attack receives an ACK from the victim, while scanning on a closed port is directly reset. Therefore, this feature captures the distinct patterns between the two attack scenarios.
Overall, it demonstrates that the interpretation provided by Genos is aligned with expert knowledge.


\begin{table}[t]
    \centering
    \scriptsize
    \renewcommand{\arraystretch}{.75}
    \caption{Interpretation of two scanning attack samples.}
    \begin{tabular}{ccc|ccc}
    \toprule
    \multicolumn{3}{c|}{Successful scanning} & \multicolumn{3}{c}{Unsuccessful scanning} \\
    Top Feature & Value & Rule & Top Feature & Value & Rule  \\ 
    \midrule
    dest\_port & $139$ & $> 1882$ & dest\_port & $9000$ & $\leq 1883$ \\
    duration & $0.0002$ & $> 13.082$ & duration & $0.663$ & $> 13.082$ \\
    count & $3$ & $> 116.48$ & count & $2$ & $> 116.48$ \\
    \bottomrule
    \end{tabular}
    \label{tab:port_scan}
\end{table}

\subsection{Updating Rules}
We first use training sets for model training and rule extraction, and apply extracted rules to the detection of validation sets. We then employ the Model Debugger to update the extracted rules using these false positives, and finally evaluate the updated rules on testing sets.

\begin{figure}[t]
    \centering
    \subfloat[CIC-IDS]{
    \includegraphics[width=0.47\linewidth]{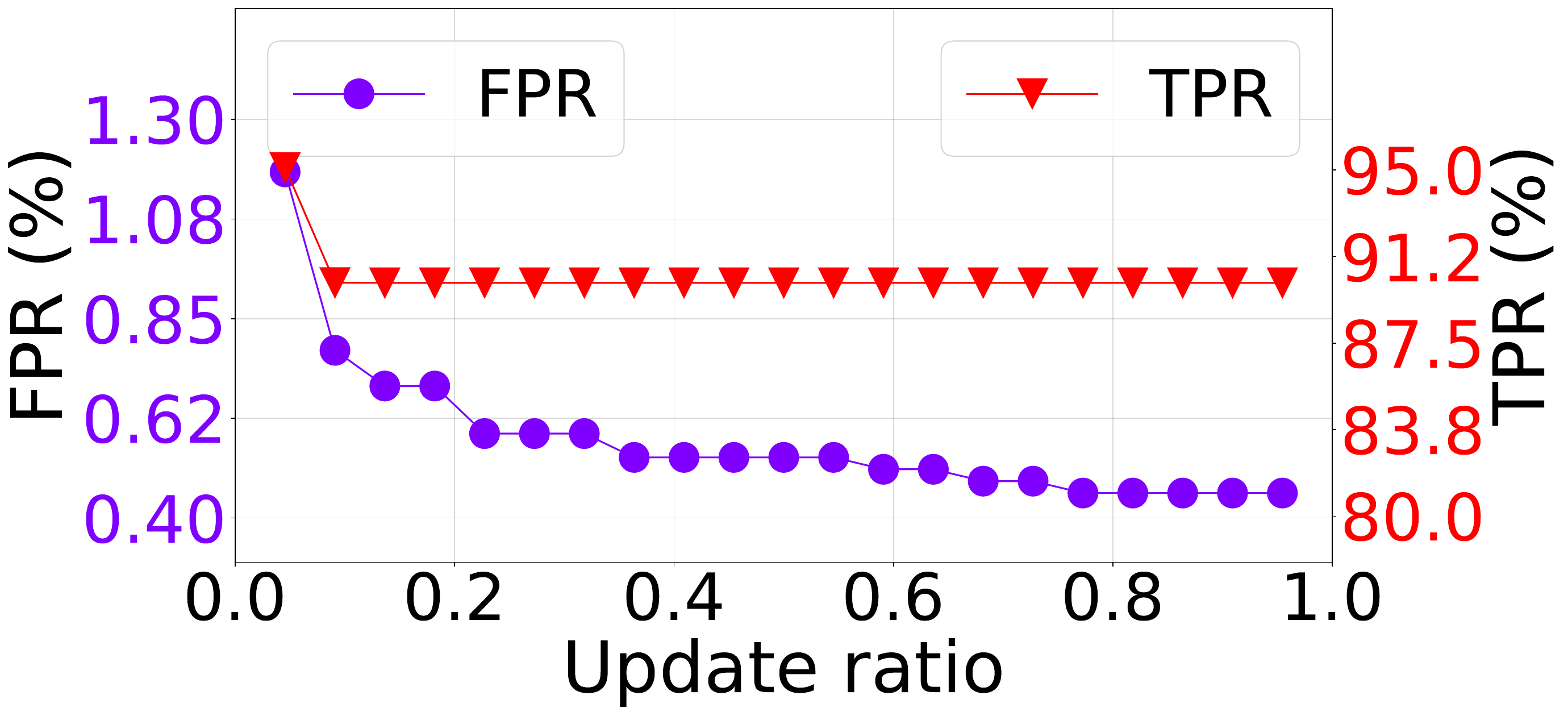}
    }
    \subfloat[TON-IoT]{
    \includegraphics[width=0.47\linewidth]{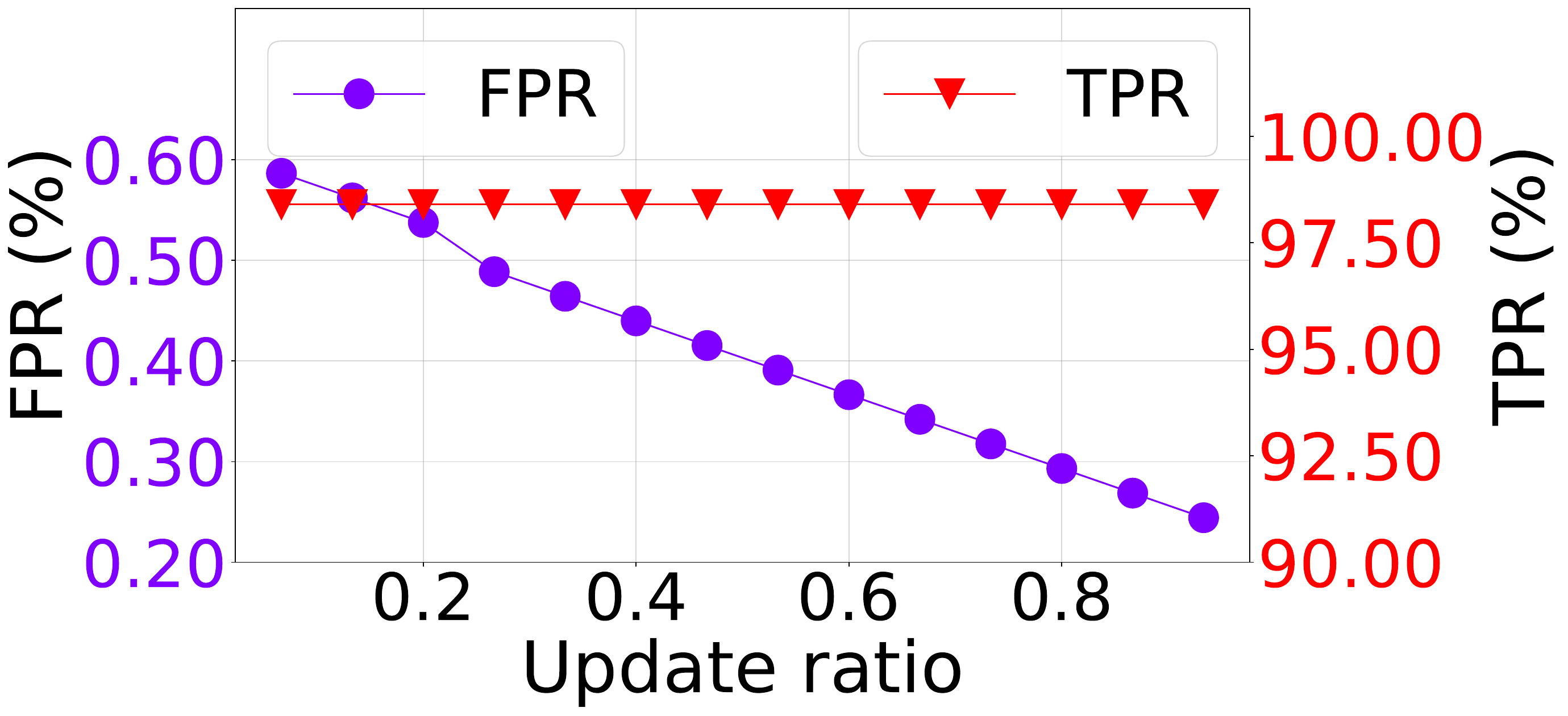}
    }    
    \caption{Updating with incremental ratios of false positives.}
    \label{fig:update_fpr}
\end{figure}

In Fig. \ref{fig:update_fpr}, as we include more false positives in rule updates, the FPR consistently decreases, showing Genos can learn from mistakes to enhance its decision-making.
Meanwhile, the TPR remains steady, indicating the persistent detection accuracy for anomalies even when rectifying false positives. This highlights the adaptability and resilience of Genos in evolving network environments, which is vital for the usability of NIDS.

\begin{figure}[t]
    \centering
    \subfloat[CIC-IDS]{
    \includegraphics[width=0.47\linewidth]{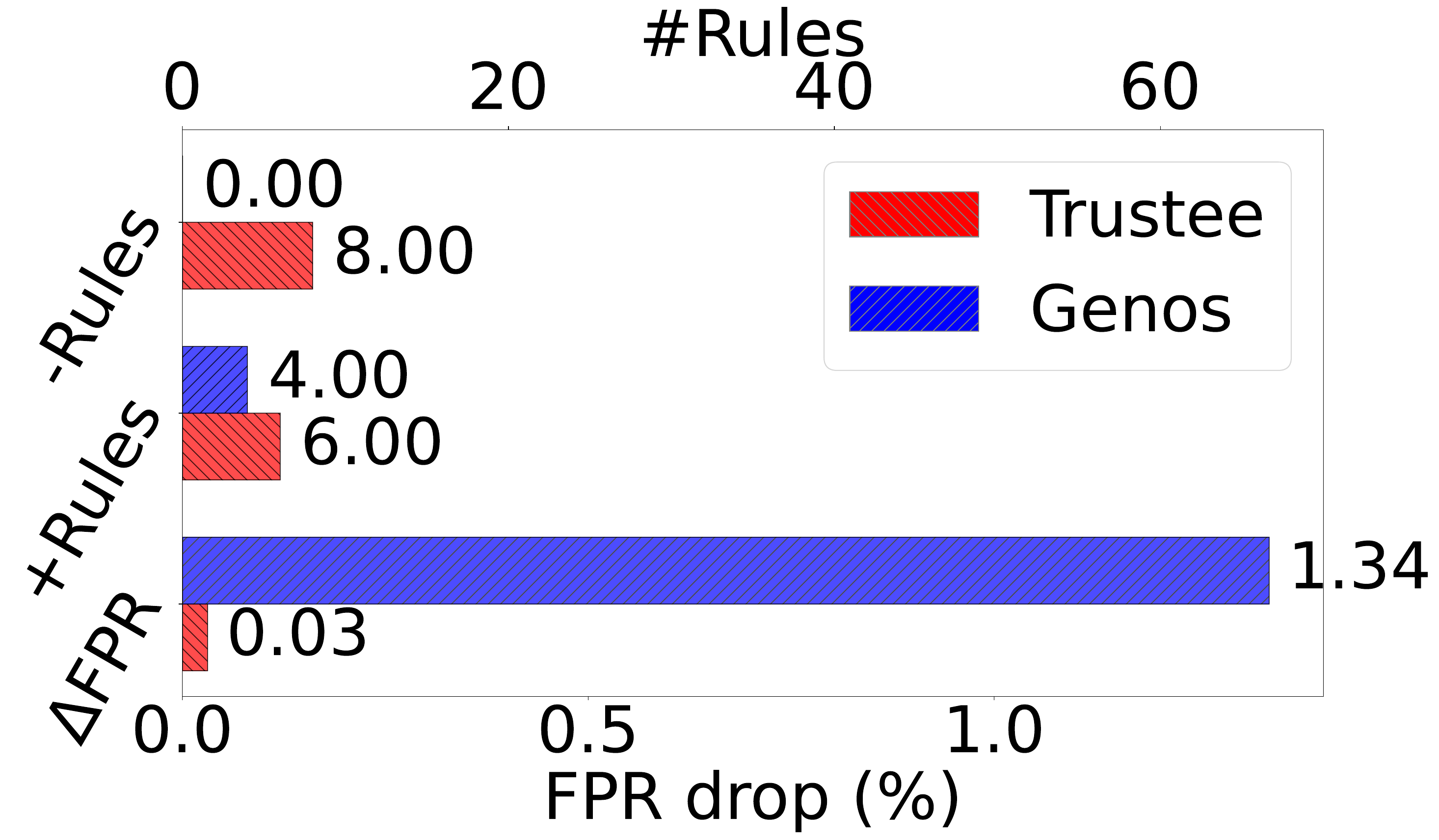}
    }
    \subfloat[TON-IoT]{
    \includegraphics[width=0.47\linewidth]{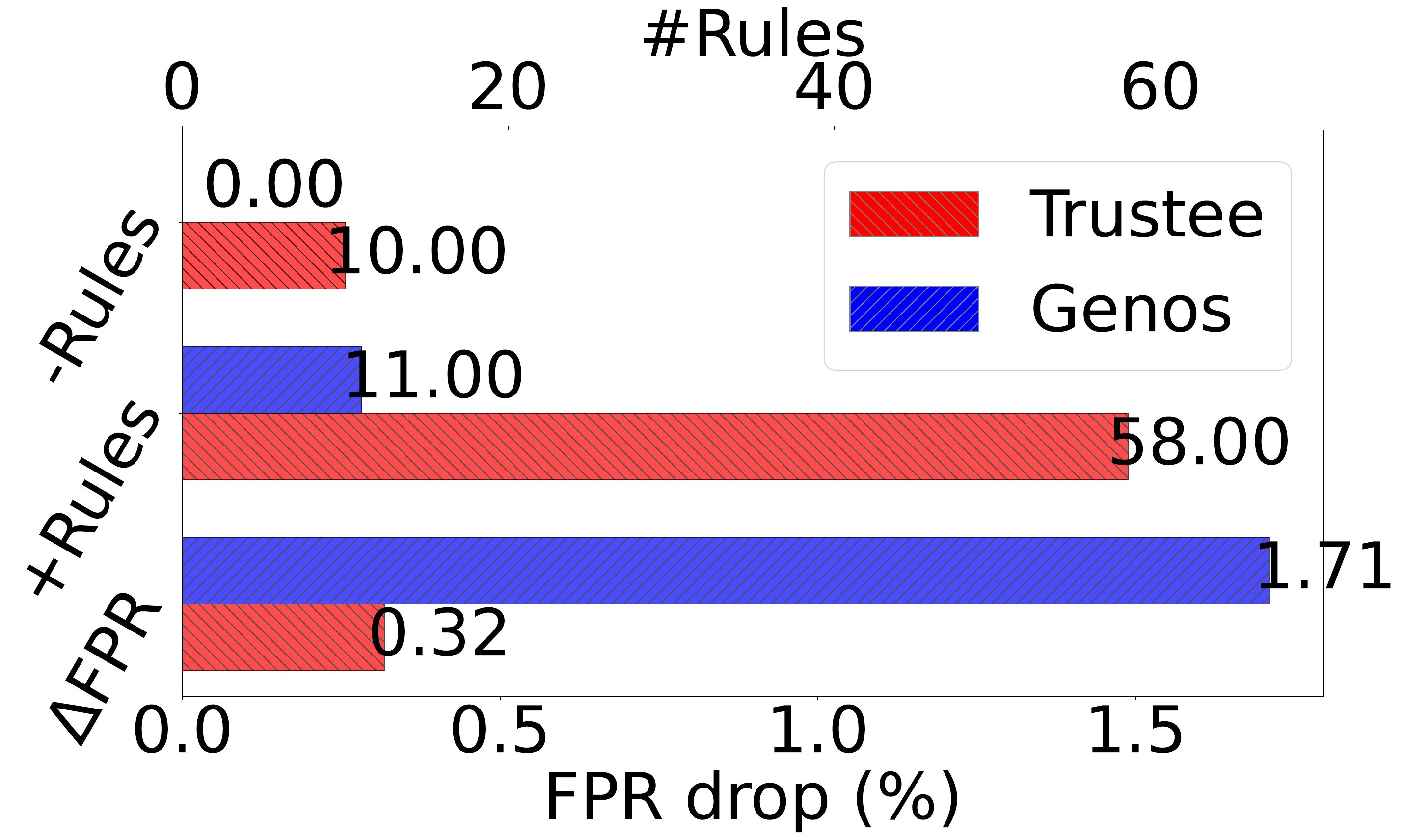}
    }    
    \caption{Change of rule numbers and FPR after updates.}
    \label{fig:rule_number}
\end{figure}

To evaluate the overhead of updates, we employ Trustee as a baseline, which is based on CART decision trees for rule extraction.
Since incremental update is not viable for Trustee, we update its rules by retraining the source model and re-extracting the rules.
Fig. \ref{fig:rule_number} shows Genos demonstrates a higher FPR drop than Trustee with fewer rule changes. Thanks to the incremental update mechanism supported by our rule extraction method, Genos can efficiently reduce the overhead of updates since only a few new rules need to be added.
In contrast, the rule set of Trustee drastically changes since the updated model changes the fundamental tree structure. 



\begin{figure}[]
    \centering
    \subfloat[Throughput]{
    \includegraphics[width=0.47\linewidth]{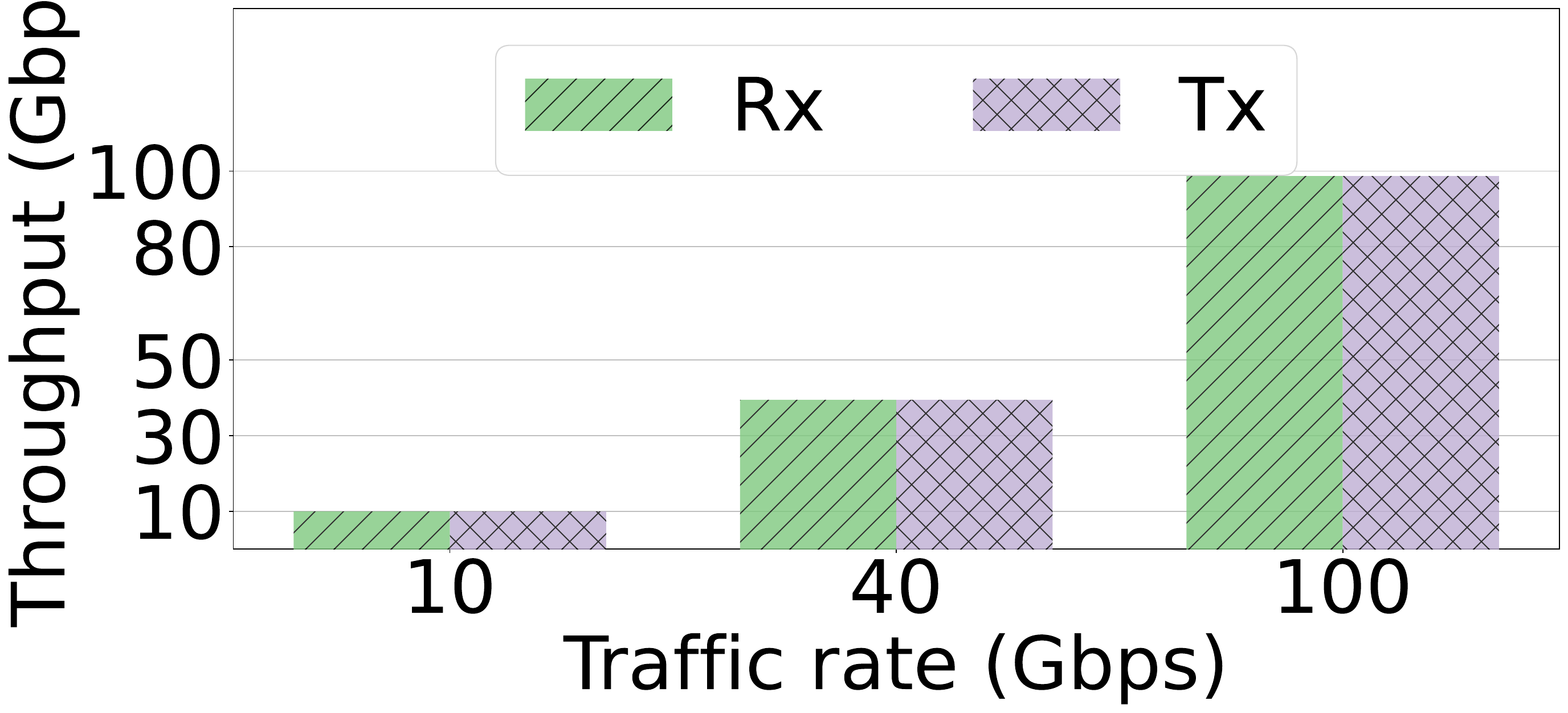}
    }
    \subfloat[Latency]{
    \includegraphics[width=0.47\linewidth]{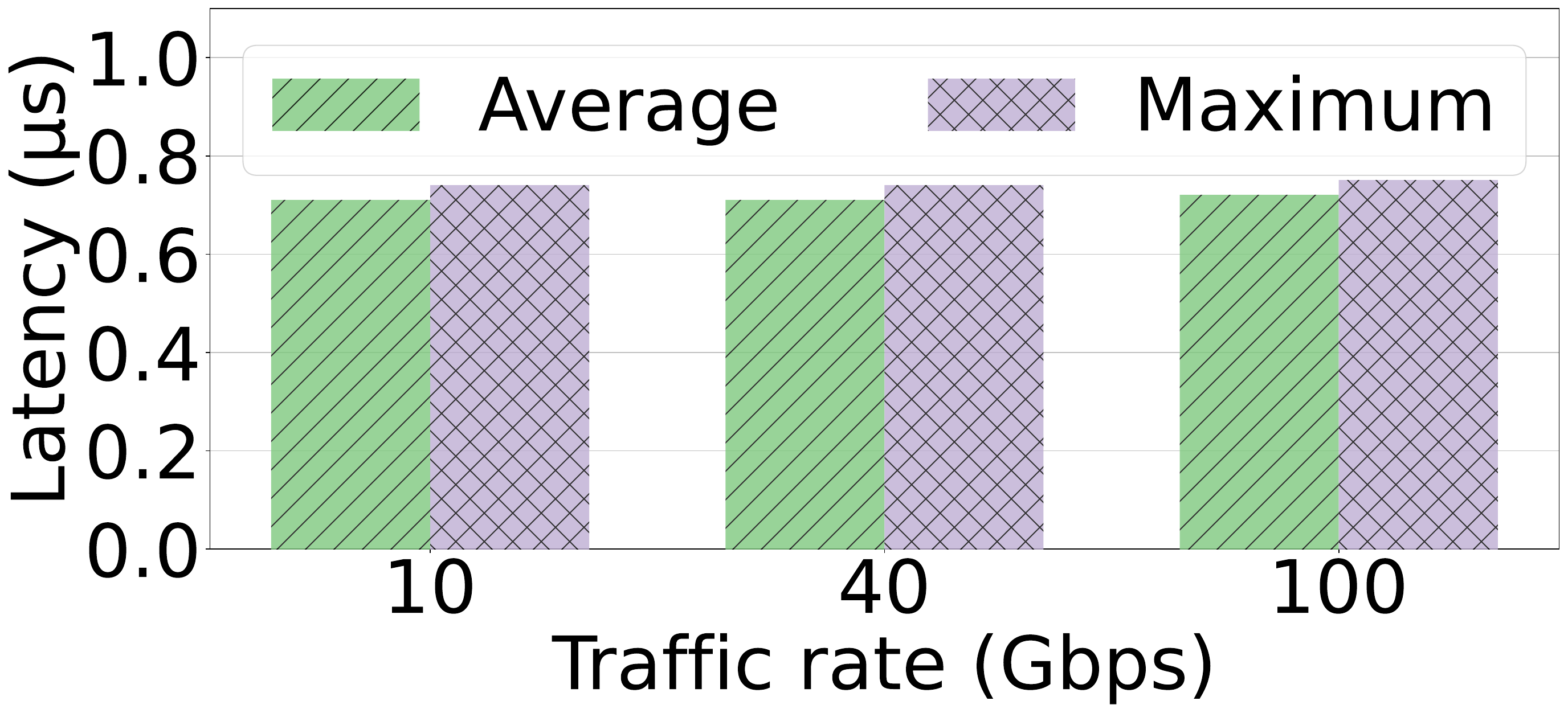}
    }    
    \caption{Runtime performance of Genos.}
    \label{fig:runtime}
\end{figure}

\subsection{Hardware Performance}

We install the P4 tables obtained by the Model Compiler and the P4 program that realizes the flow-level feature extractor and rule matching on the hardware switch. We use a traffic generator (Keysight XGS12) to generate high-speed traffic and evaluate the hardware performance with respect to throughput, processing latency, and resource consumption.

We conduct experiments with loads of 10 Gbps, 40 Gbps, and 100 Gbps. As shown in Fig. \ref{fig:runtime}, Genos exhibits exceptional performance, maintaining a perfect match between switch throughput and traffic rate. This ensures that the switch can efficiently process incoming traffic without bottlenecks or packet loss. 
Further, the processing latency is impressively low, at around 0.74 microseconds ($1\mu \text{s}$=$10^{-6}\text{s}$). 
In comparison to A-NIDS deployment on the control plane, even a highly efficient approach using DPDK \cite{whisper} can only achieve 12.65 Gbps of throughput and 0.047 seconds of latency.
The results show that Genos, through its general in-network approach, achieves high throughput and real-time online intrusion detection.

For memory resources, Genos occupies 16.9\% SRAM and 0.3\% TCAM. 
For computation resources, Genos utilizes 10 stages, 16.3\% eMatch Xbar, 0.5\% tMatch Xbar, 26.6\% Hash bit, and 52.1\% ALU. The relatively heavy utilization of the last two resources is attributed to the flow-level feature extraction mechanism.
Note that the basic packet forwarding function of switches does not use ALU extensively, so it will not be significantly affected by Genos.
Overall, Genos is efficient and practical for deployment on switch ASICs.


\section{Related Work}
\subsection{Programmable Data Plane}
There have been other works focusing on offloading various network tasks on programmable data planes, such as measurement \cite{measurement1,measurement2,measurement3}, load balancing \cite{lb}, RTT monitoring \cite{rtt}, failure detection \cite{failure}, and DDoS mitigation \cite{ddos1,ddos2}.
The primary distinction between these methods and ours lies in the approach they take to perform the tasks. While the aforementioned works mainly depend on threshold-driven logic, our framework utilizes ML/DL models to achieve in-network A-NIDS to reduce the reliance on hand-crafted thresholds.

\subsection{Model Extraction and Explanation}
There are other related works in the field of model extraction and explanation. However, these methods are either model/architecture-specific \cite{model-spec1,model-spec2}, local explanations that generate a rule for a single data sample \cite{local1,local2,local3_sec,local4_sec}, or extracting rules that need additional calculation (e.g., linear equations) \cite{global1,global2}. Our method is different from theirs as we aim to generate model-agnostic, global, and axis-aligned simple rules in an unsupervised manner, which are easily translatable into P4 tables for efficient deployment on switching ASICs.

\section{Conclusion}
This paper proposes Genos, a framework for the general in-network deployment of A-NIDS models, which can realize unsupervised model extraction and translation, produce explanations for predictions, and incrementally update deployed rules to handle false positives.
Extensive experiments show that Genos can accurately detect various malicious traffic and reach line-rate throughput for online intrusion detection.
One limitation of Genos is that it may not be suitable for handling raw data representations as input (e.g., raw packet bytes used in \cite{dark} that entail additional convolutional layers for latent feature extraction), as Genos treats every dimension as a semantic feature. However, such representations are criticized by \cite{nprint} for the possible misalignment of header fields for different protocols, decreasing model performance.
Nonetheless, we leave the exploration of interpreting models using other types of data representations as our future work to promote versatility.
The authors have provided public access to their code and/or data at \url{https://github.com/Ruoyu-Li/Genos-INFOCOM24}.

\section*{Acknowledgement}
This work is supported by National Key Research and Development Program of China under grant No. 2022YFB3105000, Major Key Project of PCL under grant No. PCL2023A06, and the Shenzhen Key Lab of Software Defined Networking under grant No. ZDSYS20140509172959989.

\bibliographystyle{IEEEtran}
\bibliography{ref}

\begin{thebibliography}{10}
\providecommand{\url}[1]{#1}
\csname url@samestyle\endcsname
\providecommand{\newblock}{\relax}
\providecommand{\bibinfo}[2]{#2}
\providecommand{\BIBentrySTDinterwordspacing}{\spaceskip=0pt\relax}
\providecommand{\BIBentryALTinterwordstretchfactor}{4}
\providecommand{\BIBentryALTinterwordspacing}{\spaceskip=\fontdimen2\font plus
\BIBentryALTinterwordstretchfactor\fontdimen3\font minus
  \fontdimen4\font\relax}
\providecommand{\BIBforeignlanguage}[2]{{%
\expandafter\ifx\csname l@#1\endcsname\relax
\typeout{** WARNING: IEEEtran.bst: No hyphenation pattern has been}%
\typeout{** loaded for the language `#1'. Using the pattern for}%
\typeout{** the default language instead.}%
\else
\language=\csname l@#1\endcsname
\fi
#2}}
\providecommand{\BIBdecl}{\relax}
\BIBdecl

\bibitem{kitsune}
Y.~Mirsky, T.~Doitshman \emph{et~al.}, ``Kitsune: An ensemble of autoencoders
  for online network intrusion detection,'' in \emph{25th Annual Network and
  Distributed System Security Symposium (NDSS)}, 2018.

\bibitem{unlearn}
M.~Du, Z.~Chen \emph{et~al.}, ``Lifelong anomaly detection through
  unlearning,'' in \emph{{ACM} {SIGSAC} Conference on Computer and
  Communications Security (CCS)}, 2019.

\bibitem{zerowall}
R.~Tang, Z.~Yang \emph{et~al.}, ``Zerowall: Detecting zero-day web attacks
  through encoder-decoder recurrent neural networks,'' in \emph{IEEE Conference
  on Computer Communications (INFOCOM)}, 2020.

\bibitem{iotargos}
Y.~Wan, K.~Xu \emph{et~al.}, ``Iotargos: {A} multi-layer security monitoring
  system for internet-of-things in smart homes,'' in \emph{{IEEE} Conference on
  Computer Communications (INFOCOM)}, 2020.

\bibitem{whisper}
C.~Fu, Q.~Li \emph{et~al.}, ``Realtime robust malicious traffic detection via
  frequency domain analysis,'' in \emph{{ACM} {SIGSAC} Conference on Computer
  and Communications Security (CCS)}, 2021.

\bibitem{ardiot}
R.~Li, Q.~Li \emph{et~al.}, ``Adriot: An edge-assisted anomaly detection
  framework against iot-based network attacks,'' \emph{IEEE Internet of Things
  J.}, vol.~9, no.~13, pp. 10\,576--10\,587, 2022.

\bibitem{iotensemble}
------, ``Iotensemble: Detection of botnet attacks on internet of things,'' in
  \emph{27th European Symposium on Research in Computer Security ({ESORICS})},
  2022.

\bibitem{hypervision}
C.~Fu, Q.~Li, and K.~Xu, ``Detecting unknown encrypted malicious traffic in
  real time via flow interaction graph analysis,'' in \emph{30th Annual Network
  and Distributed System Security Symposium (NDSS)}, 2023.

\bibitem{net2net}
G.~Siracusano and R.~Bifulco, ``In-network neural networks,'' \emph{CoRR}, vol.
  abs/1801.05731, 2018.

\bibitem{bnn_ids}
Q.~Qin, K.~Poularakis \emph{et~al.}, ``Line-speed and scalable intrusion
  detection at the network edge via federated learning,'' in \emph{2020 {IFIP}
  Networking Conference}, 2020.

\bibitem{jointNIDS}
T.~Dao and H.~Lee, ``Jointnids: Efficient joint traffic management for
  on-device network intrusion detection,'' \emph{{IEEE} Trans. Veh. Technol.},
  vol.~71, no.~12, pp. 13\,254--13\,265, 2022.

\bibitem{iisy}
Z.~Xiong and N.~Zilberman, ``Do switches dream of machine learning?: Toward
  in-network classification,'' in \emph{18th {ACM} Workshop on Hot Topics in
  Networks (HotNets)}, 2019.

\bibitem{planter2}
C.~Zheng, M.~Zang \emph{et~al.}, ``Automating in-network machine learning,''
  \emph{CoRR}, vol. abs/2205.08824, 2022.

\bibitem{IN_classification}
B.~M. Xavier, R.~S. Guimaraes \emph{et~al.}, ``Programmable switches for
  in-networking classification,'' in \emph{{IEEE} Conference on Computer
  Communications (INFOCOM)}, 2021.

\bibitem{pforest}
C.~Busse{-}Grawitz, R.~Meier \emph{et~al.}, ``pforest: In-network inference
  with random forests,'' \emph{CoRR}, vol. abs/1909.05680, 2019.

\bibitem{planter}
C.~Zheng and N.~Zilberman, ``Planter: seeding trees within switches,'' in
  \emph{{ACM} {SIGCOMM} Conference, Poster and Demo Sessions}, 2021.

\bibitem{mousika}
G.~Xie, Q.~Li \emph{et~al.}, ``Mousika: Enable general in-network intelligence
  in programmable switches by knowledge distillation,'' in \emph{{IEEE}
  Conference on Computer Communications (INFOCOM)}, 2022.

\bibitem{redt}
Y.~Li, J.~Bai \emph{et~al.}, ``Rectified decision trees: Exploring the
  landscape of interpretable and effective machine learning,'' \emph{CoRR},
  vol. abs/2008.09413, 2020.

\bibitem{trustee}
A.~S. Jacobs, R.~Beltiukov \emph{et~al.}, ``Ai/ml for network security: The
  emperor has no clothes,'' in \emph{ACM SIGSAC Conference on Computer and
  Communications Security (CCS)}, 2022.

\bibitem{blackbox_extract}
O.~Bastani, C.~Kim, and H.~Bastani, ``Interpreting blackbox models via model
  extraction,'' \emph{CoRR}, vol. abs/1705.08504, 2017.

\bibitem{ocsvm_ids}
A.~Binbusayyis and T.~Vaiyapuri, ``Unsupervised deep learning approach for
  network intrusion detection combining convolutional autoencoder and one-class
  {SVM},'' \emph{Appl. Intell.}, vol.~51, no.~10, pp. 7094--7108, 2021.

\bibitem{horuseye}
Y.~Dong, Q.~Li \emph{et~al.}, ``Horuseye: Realtime iot malicious traffic
  detection framework with programmable switches,'' in \emph{32nd {USENIX}
  Security Symposium ({USENIX} Security)}, 2023.

\bibitem{p4}
P.~Bosshart, D.~Daly \emph{et~al.}, ``{P4:} programming protocol-independent
  packet processors,'' \emph{Comput. Commun. Rev.}, vol.~44, no.~3, pp. 87--95,
  2014.

\bibitem{decision_list}
B.~Letham, C.~Rudin \emph{et~al.}, ``Interpretable classifiers using rules and
  bayesian analysis: Building a better stroke prediction model,'' \emph{CoRR},
  vol. abs/1511.01644, 2015.

\bibitem{softdt}
N.~Frosst and G.~E. Hinton, ``Distilling a neural network into a soft decision
  tree,'' \emph{CoRR}, vol. abs/1711.09784, 2017.

\bibitem{netbeacon}
G.~Zhou, Z.~Liu \emph{et~al.}, ``An efficient design of intelligent network
  data plane,'' in \emph{32st {USENIX} Security Symposium ({USENIX Security})},
  2023.

\bibitem{lime}
M.~T. Ribeiro, S.~Singh, and C.~Guestrin, ``"why should i trust you?":
  Explaining the predictions of any classifier,'' in \emph{22nd ACM SIGKDD
  International Conference on Knowledge Discovery and Data Mining (KDD)}, 2016.

\bibitem{cart}
L.~Breiman, J.~H. Friedman \emph{et~al.}, \emph{Classification and Regression
  Trees}.\hskip 1em plus 0.5em minus 0.4em\relax Wadsworth, 1984.

\bibitem{fsgm}
I.~J. Goodfellow, J.~Shlens, and C.~Szegedy, ``Explaining and harnessing
  adversarial examples,'' in \emph{3rd International Conference on Learning
  Representations (ICLR)}, 2015.

\bibitem{local1}
S.~M. Lundberg and S.~Lee, ``A unified approach to interpreting model
  predictions,'' in \emph{Advances in Neural Information Processing Systems 30:
  Annual Conference on Neural Information Processing Systems}, 2017.

\bibitem{local2}
M.~T. Ribeiro, S.~Singh, and C.~Guestrin, ``Anchors: High-precision
  model-agnostic explanations,'' in \emph{{AAAI} Conference on Artificial
  Intelligence (AAAI)}, 2018.

\bibitem{local3_sec}
W.~Guo, D.~Mu \emph{et~al.}, ``Lemna: Explaining deep learning based security
  applications,'' in \emph{ACM SIGSAC Conference on Computer and Communications
  Security (CCS)}, 2018.

\bibitem{local4_sec}
D.~Han, Z.~Wang \emph{et~al.}, ``Deepaid: Interpreting and improving deep
  learning-based anomaly detection in security applications,'' in \emph{2021
  ACM SIGSAC Conference on Computer and Communications Security (CCS)}, 2021.

\bibitem{cicflowmeter}
``Cicflowmeter,'' \url{https://github.com/ahlashkari/CICFlowMeter}, Canadian
  Institute for Cybersecurity, 2016.

\bibitem{netflow}
``Netflow,''
  \url{https://cisco.com/c/dam/en/us/td/docs/security/stealthwatch/netflow/Cisco_NetFlow_Configuration.pdf},
  Cisco, 2023.

\bibitem{cicids}
I.~Sharafaldin, A.~H. Lashkari, and A.~A. Ghorbani, ``Toward generating a new
  intrusion detection dataset and intrusion traffic characterization,'' in
  \emph{4th International Conference on Information Systems Security and
  Privacy (ICISSP)}, 2018.

\bibitem{ton-iot}
T.~M. Booij, I.~Chiscop \emph{et~al.}, ``Ton{\_}iot: The role of heterogeneity
  and the need for standardization of features and attack types in iot network
  intrusion data sets,'' \emph{{IEEE} Internet Things J.}, vol.~9, no.~1, pp.
  485--496, 2022.

\bibitem{measurement1}
S.~Wang, C.~Sun \emph{et~al.}, ``Martini: Bridging the gap between network
  measurement and control using switching asics,'' in \emph{28th {IEEE}
  International Conference on Network Protocols (ICNP)}, 2020.

\bibitem{measurement2}
J.~Xing, Q.~Kang, and A.~Chen, ``Netwarden: Mitigating network covert channels
  while preserving performance,'' in \emph{29th {USENIX} Security Symposium
  ({USENIX} Security)}, 2020.

\bibitem{measurement3}
D.~Barradas, N.~Santos \emph{et~al.}, ``Flowlens: Enabling efficient flow
  classification for ml-based network security applications,'' in \emph{28th
  Annual Network and Distributed System Security Symposium (NDSS)}, 2021.

\bibitem{lb}
R.~Miao, H.~Zeng \emph{et~al.}, ``Silkroad: Making stateful layer-4 load
  balancing fast and cheap using switching asics,'' in \emph{{ACM} {SIGCOMM}
  Conference (SIGCOMM)}, 2017.

\bibitem{rtt}
S.~Sengupta, H.~Kim, and J.~Rexford, ``Continuous in-network round-trip time
  monitoring,'' in \emph{{ACM} {SIGCOMM} Conference (SIGCOMM)}, 2022.

\bibitem{failure}
E.~C. Molero, S.~Vissicchio, and L.~Vanbever, ``Fast in-network \emph{GraY}
  failure detection for isps,'' in \emph{{ACM} {SIGCOMM} Conference (SIGCOMM)},
  2022.

\bibitem{ddos1}
Z.~Liu, H.~Namkung \emph{et~al.}, ``Jaqen: {A} high-performance switch-native
  approach for detecting and mitigating volumetric ddos attacks with
  programmable switches,'' in \emph{30th {USENIX} Security Symposium ({USENIX}
  Security)}, 2021.

\bibitem{ddos2}
M.~Zhang, G.~Li \emph{et~al.}, ``Poseidon: Mitigating volumetric ddos attacks
  with programmable switches,'' in \emph{27th Annual Network and Distributed
  System Security Symposium ({NDSS})}, 2020.

\bibitem{model-spec1}
D.~Kazhdan, B.~Dimanov \emph{et~al.}, ``{MEME:} generating {RNN} model
  explanations via model extraction,'' \emph{CoRR}, vol. abs/2012.06954, 2020.

\bibitem{model-spec2}
P.~Liznerski, L.~Ruff \emph{et~al.}, ``Explainable deep one-class
  classification,'' in \emph{9th International Conference on Learning
  Representations ({ICLR})}, 2021.

\bibitem{global1}
M.~W. Craven and J.~W. Shavlik, ``Using sampling and queries to extract rules
  from trained neural networks,'' in \emph{International Conference on Machine
  Learning (ICML)}, 1994.

\bibitem{global2}
Z.~Zhou, Y.~Jiang, and S.~Chen, ``Extracting symbolic rules from trained neural
  network ensembles,'' \emph{{AI} Commun.}, vol.~16, no.~1, pp. 3--15, 2003.

\bibitem{dark}
G.~Mar{\'{\i}}n, P.~Casas, and G.~Capdehourat, ``Deep in the dark - deep
  learning-based malware traffic detection without expert knowledge,'' in
  \emph{{IEEE} Security and Privacy Workshops (SPW)}, 2019.

\bibitem{nprint}
J.~Holland, P.~Schmitt \emph{et~al.}, ``New directions in automated traffic
  analysis,'' in \emph{{ACM} {SIGSAC} Conference on Computer and Communications
  Security (CCS)}, 2021.

\end{thebibliography}
\end{document}